\documentclass[10pt,letterpaper]{article}
\usepackage[top=0.85in,left=2.75in,footskip=0.75in]{geometry}

\usepackage{amsmath,amssymb}
\usepackage{changepage}
\usepackage[utf8x]{inputenc}
\usepackage{textcomp,marvosym}
\usepackage{cite}
\usepackage{nameref,hyperref}
\usepackage{microtype}
\DisableLigatures[f]{encoding = *, family = * }
\usepackage[table]{xcolor}
\usepackage{array}
\usepackage[final]{pdfpages}

\newcolumntype{+}{!{\vrule width 2pt}}

\newlength\savedwidth

\raggedright
\usepackage[aboveskip=1pt,labelfont=bf,labelsep=period,justification=raggedright,singlelinecheck=off]{caption}

\makeatletter
\renewcommand{\@biblabel}[1]{\quad#1.}
\makeatother

\usepackage{lastpage,fancyhdr,graphicx}
\usepackage{epstopdf}

\fancyheadoffset[L]{2.25in}
\fancyfootoffset[L]{2.25in}
\newcommand{\smin}{s_{\text{min}}}
\newcommand{\smax}{s_{\text{max}}}
\newcommand{\tsmin}{\tilde{s}_{\text{min}}}

\begin{document}
\vspace*{0.2in}

\begin{flushleft}
{\Large
\textbf\newline{Association between population distribution and urban GDP scaling}
}
\newline
\\
Haroldo\ V.\ Ribeiro\textsuperscript{1,*},
Milena Oehlers\textsuperscript{2},
Ana~I.~Moreno-Monroy\textsuperscript{3},
J\"urgen P.\ Kropp\textsuperscript{2,4},
Diego Rybski\textsuperscript{2,5,$\dagger$},
\\
\bigskip
\textbf{1} Departamento de F\'isica, Universidade Estadual de Maring\'a -- Maring\'a, PR 87020-900, Brazil
\\
\textbf{2} Potsdam Institute for Climate Impact Research -- PIK, Member of Leibniz Association, P.O.\ Box 601203, 14412 Potsdam, Germany
\\
\textbf{3} OECD, 1 Rue Andr\'e Pascal, 75016, Paris, France; Department of Geography and Planning, University of Liverpool, Chatham St, Liverpool L69 7ZT, United Kingdom
\\
\textbf{4} Institute for Environmental Science and Geography, University of Potsdam, 14476 Potsdam, Germany
\\
\textbf{5} Department of Environmental Science Policy and Management, University of California Berkeley, 130 Mulford Hall \#3114, Berkeley, CA 94720, USA
\\
\bigskip

* hvr@dfi.uem.br, $\dagger$ ca-dr@rybski.de

\end{flushleft}
\section*{Abstract}
Urban scaling and Zipf's law are two fundamental paradigms for the science of cities. These laws have mostly been investigated independently and are often perceived as disassociated matters. Here we present a large scale investigation about the connection between these two laws using population and GDP data from almost five thousand consistently-defined cities in 96 countries. We empirically demonstrate that both laws are tied to each other and derive an expression relating the urban scaling and Zipf exponents. This expression captures the average tendency of the empirical relation between both exponents, and simulations yield very similar results to the real data after accounting for random variations. We find that while the vast majority of countries exhibit increasing returns to scale of urban GDP, this effect is less pronounced in countries with fewer small cities and more metropolises (small Zipf exponent) than in countries with a more uneven number of small and large cities (large Zipf exponent). Our research puts forward the idea that urban scaling does not solely emerge from intra-city processes, as population distribution and scaling of urban GDP are correlated to each other.

\section*{Introduction}

Physical measurements such as weight or size of objects are always confined to specific scales. However, the outcomes of several natural phenomena and socio-economic processes can extend across multiple orders of magnitude~\cite{westscale2017}. Examples of such systems are as diverse as earthquakes~\cite{gutenberg1944frequency}, stock market fluctuations~\cite{buchaudpower2001}, casualties in human insurgencies~\cite{bohorquezcommon2009}, and fracture of materials~\cite{ribeiroanalogies2015}. These systems usually share non-trivial statistical regularities manifested in the form of power-law distributions or power-law relations, the so-called scaling laws. Cities are among these systems as they occur in sizes from thousands to tens of million inhabitants. Urban systems are also well known to follow scaling relations in time and space, as summarized by Batty~\cite{battynew2013} in the ``seven laws of urban scaling''.

Zipf's law~\cite{zipfhuman2012, auerbacghesetz1913} and Bettencourt-West law~\cite{bettencourtgrowth2007} are two of the best-known scaling laws emerging in urban systems. Initially observed by Auerbach~\cite{auerbacghesetz1913} and then popularized by Zipf~\cite{zipfhuman2012}, Zipf's law for cities states that the distribution of city populations, for a given region or country, is approximated by an inverse power-law function (or the rank-size rule), implying that there are plenty of small cities and very few metropolises. Bettencourt-West law~\cite{bettencourtgrowth2007}, better known as ``urban scaling'', establishes a power-law relation between urban indicators and city population. Urban scaling is well illustrated by the concept of agglomeration economies in cities~\cite{marshallprinciples1890,sveikauskasproductivity1975}, in which the power-law association between urban wealth and population implies that urban wealth increases more than proportionally with city population.

Fostered by a recent deluge of highly-detailed city data, researchers from diverse disciplines such as geography, economics, or physics have devoted ongoing efforts in identifying and understanding fundamental principles and regularities underlying urban systems~\cite{battynew2013,westscale2017,barthelemy2016structure,barthelemy2019statistical,lobo2020urban}. While most of these works are either concerned with Zipf's law~\cite{gabaix1999zipf,gabaix2004evolution,nitschzipf2005,soo2005zipf,batty2006rank,berrycity2012,schmidheiny2015pan,li2015allometric,cottineaumetamipf2017,zhang2019scale,mori2020common} or urban scaling~\cite{bettencourt2013origins,strano2016rich,gomez2018explaining,depersin2018global,ribeiro2017model,keuschnigg2019scaling,pumain2019two,balland2020complex,bettencourtinterpretation2020}, very few have tackled the relationship between both. Zipf's law and urban scaling have mostly been studied independently because it is commonly assumed that both laws are independent descriptions of urban systems across countries. The few investigations on possible connections between Zipf's law and urban scaling have a more local character and explore this association within countries~\cite{gomezstatistics2012,alvesempirical2014}. The work of Gomez-Lievano \textit{et al.}~\cite{gomezstatistics2012} is seminal in this regard and has shown that (under certain conditions) urban scaling can be related to Zipf's law when urban indicators are also power-law distributed. This connection is obtained from probability distributions and shows that the urban scaling exponent has an upper limit determined by both the Zipf exponent and the power-law exponent that characterizes the urban indicator distribution. This result however does not imply that these three exponents are directly associated with each other. In fact, random permutations of population and urban indicator values only have the effect of removing a possible correlation between the two variables, but do not have the effect of changing their distribution.

To date, there has been no attempt to empirically investigate the association between urban scaling and Zipf's law across a large number of countries. The paucity of such studies certainly reflects the lack of consistent and comparable data across countries. A convincing comparison between these two laws demands a unified city definition across the globe as well as measures of population and urban indicators based on this same definition. And it was not until very recently that satisfying this requirement has become possible. Here we use a new harmonized city definition for investigating Zipf's law and urban scaling over almost five thousand cities in 96 countries. By analyzing population and gross domestic product (GDP) of these cities per country, we estimate the Zipf and the urban scaling exponents to probe for a possible relation between these two scaling laws. Our empirical results show that both exponents are indeed related to each other and that a functional form of this association can be exactly derived from scaling relations emerging at the country level.

We demonstrate that Zipf's law and urban scaling imply a power-law relation between total urban population and total urban GDP of countries, where the country scaling exponent is dependent on the Zipf and the urban scaling exponents. Because country-aggregated values of urban population and GDP are fixed, there is only one country scaling exponent for total GDP, which in turn associates the urban scaling exponent to the Zipf exponent. We verify the integrity of our model by estimating the country-scaling exponents from the empirical relationship between the Zipf and the urban scaling exponents, and also by showing that numerical simulations yield results very similar to those obtained from real-world data. The connection between both exponents shows that urban scaling does not only emerge from processes occurring within the city boundaries; instead, it suggests that the population distribution of an urban system does affect urban scaling or vice versa. For the particular case of urban GDP, while almost all countries exhibit increasing returns to scale (urban scaling exponent greater than one), our findings indicate this effect is smaller in countries with a more balanced number of small and large cities (small Zipf exponent) than in countries with a more unbalanced number of small and large cities (large Zipf exponent).

\section*{Results}

\subsection*{Empirical connection between both scaling laws}

We start by revisiting Zipf's law and the urban scaling law. Zipf's law for cities~\cite{zipfhuman2012} establishes that the rank $r$ and the city population $s$ of an urban system are related through a power-law function
\begin{equation}\label{eq:zipf}
    r \sim s^{-\alpha}\,,
\end{equation}
where $\alpha>0$ is the Zipf exponent. The estimated $\alpha$ varies across countries and epochs, but $\alpha\approx 1$ is typically found in empirical studies~\cite{soo2005zipf,cottineaumetamipf2017}. Urban scaling laws~\cite{bettencourtgrowth2007} commonly refer to power-law associations between an urban indicator $y$ and the city population $s$ within an urban system, that is,
\begin{equation}\label{eq:scaling}
y = c\,s^{\beta}\,,
\end{equation}
where $c>0$ is a constant and $\beta$ is the urban scaling exponent. The value of $\beta$ depends on the type of urban indicator, but increasing returns to scale ($\beta>1$) are usually reported for socio-economic indicators, decreasing returns to scale ($\beta<1$) for infrastructure indicators, and constant returns to scale ($\beta=1$) for indicators related to individual needs~\cite{bettencourtgrowth2007}. The values of $\alpha$ and $\beta$ are also susceptible to different city definitions~\cite{arcauteconstructing2014,cottineaumetamipf2017}, and we thus need a unified definition across different urban systems to test a possible association between their values.

To do so, we use a generalized definition of functional urban areas (FUA) recently proposed by Moreno-Monroy \textit{et al.}~\cite{ghslfua2019schiavina,moreno2020metropolitan}. The concept of FUA was initially developed for countries of the OECD and Europe as a unified definition of metropolitan areas, consisting of high-density urban cores and their surrounding areas of influence or commuting areas. The generalized definition we use (so-called eFUA or GHSL-FUA) represents an extension of this concept to countries of the entire globe. By considering the areas of influence of urban cores, eFUAs give a less fragmented representation of the city size distribution because dense clusters proximate to urban cores are rightly not considered as independent cities. While eFUAs delineate world-wide comparable city boundaries, the majority of urban indicators are available at the level of local administrative units, that in addition to not being centralized into a global data set, may change from country to country. We avoid this problem by considering a global gridded data set for GDP~\cite{kummu2018gridded} (see Materials and Methods). Thus, by combining these two data sources (see Materials and Methods and Fig.~1 in S1 Appendix), we create a consistent data set comprising the population ($s$) and the GDP ($y$) of 4,571 cities from 96 countries.

\begin{figure*}[!ht]
\begin{adjustwidth}{-2.25in}{0in}
\begin{center}
\includegraphics[width=1.4\textwidth,keepaspectratio]{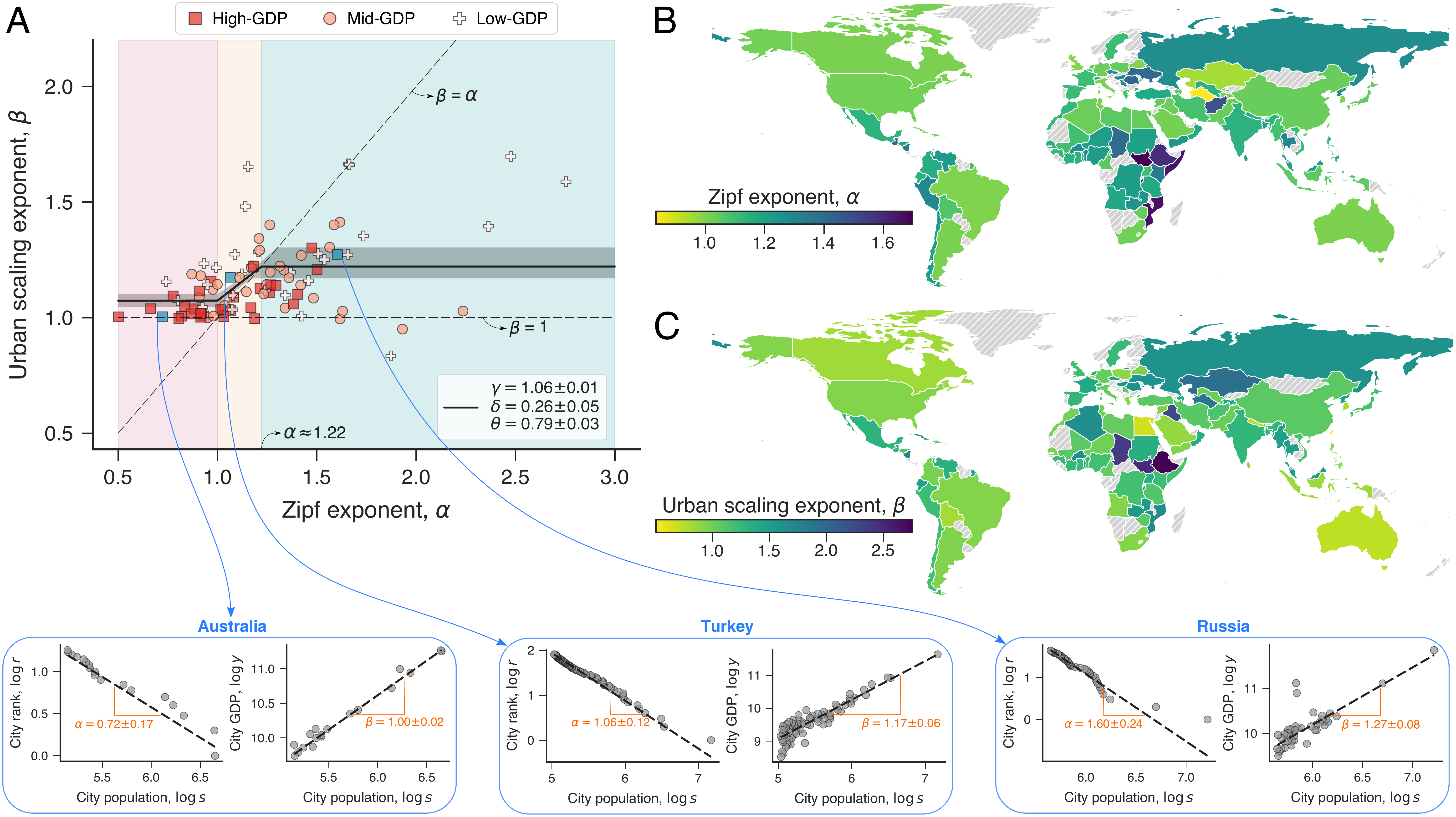}
\end{center}
\caption{{Association between the urban scaling exponent $\beta$ and the Zipf exponent $\alpha$.} (A) Values of $\beta$ versus $\alpha$ for each country in our data set. The three different markers distinguish countries according to the tercile values of the total urban GDP distribution (for instance, high-GDP countries have highest $\approx33$\% GDP values). The horizontal dashed line shows the $\beta=1$ while the inclined dashed line represents the $\beta=\alpha$ relationship. As we shall discuss in the next section, the continuous line shows the model of Eq.~(\ref{eq:betaF1}) adjusted to data and the gray shaded region stands for the 95\% confidence band. The colored background represents the different intervals of $\alpha$ defined in Eq.~(\ref{eq:betaF1}). The insets (indicated by blue arrows) illustrate Zipf's law for city population (left) and the urban scaling relationship of city GDP (right) for three countries (see S1 Appendix, Sec.~4 for all considered countries). We have verified that the model of Eq.~(\ref{eq:betaF1}) provides a significantly better description of data when compared with a null model where $\beta$ is independent of $\alpha$ (intercept-only model: $\beta=\text{constant}$, see Fig.~4 in S1 Appendix), and that statistical correlations between $\beta$ and $\alpha$ are significantly enhanced in the (orange shaded) region where our model predicts a linear correspondence between the exponents (Fig.~5 in S1 Appendix). Maps in panels (B) and (C) show the color-coded values of $\alpha$ and $\beta$ for each country (light gray indicates missing values).}
\label{figure1}
\end{adjustwidth}
\end{figure*}

Having this harmonized data set, we estimate the Zipf exponent $\alpha$ and urban scaling exponent $\beta$ after grouping the values of $s$ and $y$ by country (see Materials and Methods). The scatter plot in Fig.~\ref{figure1}A depicts the values of $\beta$ versus $\alpha$, where the insets show Zipf's law and urban scaling of city GDP for three countries. As these three examples illustrate, the data follows Zipf's law and urban scaling with deviations comparable with other studies about these two laws (see S1 Appendix, Sec.~4 for all countries). The world maps in Figs.~\ref{figure1}B and C indicate the regional distribution of the values of $\alpha$ and $\beta$. In line with the meta-analysis of Refs.~\cite{soo2005zipf,cottineaumetamipf2017}, we find Zipf exponents $\alpha$ roughly distributed around $1$ with average value and standard deviation equal to $1.24$ and $0.38$, respectively (Fig.~2A in S1 Appendix). In turn, $\approx94$\% of the countries exhibit urban scaling exponents with GDP ($\beta$) larger than 1 (Fig.~2B in S1 Appendix), and the average value and standard deviation of $\beta$ are $1.16$ and $0.17$, respectively. This result agrees with the idea that economic indicators display increasing returns to scale~\cite{bettencourtgrowth2007,bettencourt2010urban}.

The results shown in Fig.~\ref{figure1} suggest a positive association between the values of $\beta$ and $\alpha$, that is, an increase of the Zipf exponent $\alpha$ appears to come along with a rise in the urban scaling exponent $\beta$ (or vice versa). This behavior is also perceptible in the world maps, where we note that regional differences in the values of $\alpha$ (Fig.~\ref{figure1}B) are similar to the ones observed for $\beta$ (Fig.~\ref{figure1}C). However, these visual similarities are mainly induced by the largest countries in land area, and a more careful analysis of these maps reveals important differences, especially in the African continent and Central Asia. These differences are more evident in the scatter plot of Fig.~\ref{figure1}A, where we observe a considerable spread in the association between both exponents. We have further quantified the association between the values of $\beta$ and $\alpha$ by estimating the Spearman rank correlation and the Pearson correlation within a sliding window of size $\Delta \alpha^\prime$ centered in $\alpha^\prime$. For different values of $\Delta \alpha^\prime$, we find that the correlations peak around $\alpha^\prime\approx1.1$ and start to decrease and become not statistically significant as the sliding window moves away from this value of $\alpha^\prime$ (Fig.~3 in S1 Appendix). This analysis thus suggests that the overall association between both exponents is non-linear and that $\beta$ may approach constant values for large and small values of $\alpha$.

\subsection*{Country scaling and the association between both exponents}
To better describe the empirical connection between the urban scaling exponent and the Zipf exponent, we hypothesize that this association can be derived from the relationship between country-aggregated values of urban population ($S$) and urban GDP ($Y$). As we shall show, the combination of Zipf's and the urban scaling laws implies a power-law relation between total urban population $S$ and total urban GDP $Y$. This country scaling law is characterized by an exponent $\gamma$ that is a function of $\beta$ and $\alpha$, which in turn yields a mathematical expression for the relation between $\beta$ and $\alpha$. To do so, we start by noticing that Zipf's law implies a power-law dependence for the absolute number of cities with population~$s$, that is, $p(s)=k s^{-(\alpha+1)}$ for $\smin<s<\smax$, where $\smin$ and $\smax$ represent a lower and an upper cutoff associated to the smallest and largest city population of a particular urban system, and $k$ is a normalization constant (see S1 Appendix, Sec.~1 for details). By combining this frequency function $p(s)$ with urban scaling (Eq.~\ref{eq:scaling}), we can write the total urban GDP of a country as
\begin{equation}\label{eq:integral_Y}
    Y = \sum_{s=\smin}^{\smax} c s^\beta p(s) \approx k c \int_{\smin}^{\smax} s^{\beta-\alpha-1} ds\,,
\end{equation}
where the normalization constant $k$ is determined by
\begin{equation}\label{eq:normalization_k}
    S = \sum_{s=\smin}^{\smax} s p(s) \approx k \int_{\smin}^{\smax} s^{-\alpha} ds\,.
\end{equation}
To solve these equations, we consider that $\smin$ and $\smax$ are power-law functions of $S$
\begin{eqnarray}
    \smin &=& a S^\delta \label{eq:cntry_scaling_smin}\,,\\
    \smax &=& b S^\theta \label{eq:cntry_scaling_smax}\,,
\end{eqnarray}
where $a>0$ and $b>0$ are constants, and $\delta>0$ and $\theta>0$ are country scaling exponents. Thus, by plugging Eqs.~(\ref{eq:cntry_scaling_smin}) and (\ref{eq:cntry_scaling_smax}) into Eq.~(\ref{eq:integral_Y}), we find 
\begin{equation}\label{eq:cntry_scaling_A}
    \begin{split}
        Y &= 
        \frac{c (\alpha-1) S}{\alpha-\beta}
        \left( 
        \frac{a^\beta b^\alpha S^{\delta\beta + \theta\alpha} - a^\alpha b^\beta S^{\theta\beta+\delta\alpha}}
             {a b^\alpha S^{\delta +\theta\alpha} - a^\alpha b S^{\theta + \delta \alpha}}
        \right)\,,
    \end{split}
\end{equation}
that in the limit of $S\gg1$ yields
\begin{equation}\label{eq:cntry_scaling}
    Y = Y_0 S^\gamma\,,
\end{equation}
where $Y_0$ is a constant and $\gamma=\gamma(\alpha,\beta,\delta,\theta)$ is the country scaling exponent for total urban GDP. As detailed in S1 Appendix, Sec.~1, the exact form of $\gamma$ depends on conditions imposed on the exponents $\alpha$, $\beta$, $\delta$, and $\theta$ (cases A.1-A.8 in S1 Appendix) which in turn emerge from determining the dominant term of Eq.~(\ref{eq:cntry_scaling_A}) in the limit of $S\gg1$. Therefore, the combination of Zipf's law (Eq.~\ref{eq:zipf}) with the urban scaling (Eq.~\ref{eq:scaling}) and the country scaling relations of $\smin$ and $\smax$ (Eqs.~\ref{eq:cntry_scaling_smin} and \ref{eq:cntry_scaling_smax}) leads us to the country scaling of total urban GDP (Eq.~\ref{eq:cntry_scaling}), where the exponent $\gamma$ depends on $\alpha$, $\beta$, $\delta$ and $\theta$. 

The country scaling relations of GDP and $\smax$ have previously been empirically observed in Refs.~\cite{pumaincity1997,zhang2010allometric,rybskicities2017}. We have also verified that Eqs.~(\ref{eq:cntry_scaling}), (\ref{eq:cntry_scaling_smin}), and (\ref{eq:cntry_scaling_smax}) hold well for the country-aggregated values in our data set with $\gamma>1$ and $\delta<\theta$ (Fig.~6 in S1 Appendix). More importantly, while $\alpha$ and $\beta$ are intra-country exponents having different values for each country, $\gamma$, $\delta$ and $\theta$ are inter-country exponents and there is only one value for each across all countries. Thus, we can solve $\gamma=\gamma(\alpha,\beta,\delta,\theta)$ for $\beta$ (S1 Appendix, Sec.~1 for details) to find
\begin{equation}\label{eq:betaF1}
  \beta = \begin{cases}
  1 + \frac{\gamma-1}{\theta} & 0<\alpha\leq1\\
  \frac{\gamma+\delta-1}{\theta} + \left(1-\frac{\delta}{\theta}\right)\alpha & 1<\alpha<1+\frac{\gamma-1}{\delta}\\
  1 + \frac{\gamma-1}{\delta} & \alpha\geq 1+\frac{\gamma-1}{\delta}
  \end{cases}\,,
\end{equation}
for $\gamma>1$ and $\delta<\theta$. This piecewise relationship implies that $\beta$ is a constant up to $\alpha=1$, where it starts to increase linearly until reaching the diagonal line given by $\beta=\alpha$, and then continues as another constant. It is worth mentioning that by writing $\beta$ as a function of $\alpha$, we are not assuming any causal direction for the association between both exponents. Indeed, we could also solve $\gamma=\gamma(\alpha,\beta,\delta,\theta)$ for $\alpha$ but this yields a non-functional dependency, that is, the horizontal plateaus defined by Eq.~(\ref{eq:betaF1}) become vertical lines when writing $\alpha$ as function of $\beta$.

We have adjusted Eq.~(\ref{eq:betaF1}) to the empirical relation between $\alpha$ and $\beta$ (see Materials and Methods) and the best fitting parameters are $\gamma=1.06\pm0.01$, $\delta=0.26\pm0.05$, and $\theta=0.79\pm0.03$. The solid line in Fig.~\ref{figure1}A represents the best fit of Eq.~(\ref{eq:betaF1}) and the colored background indicates the different ranges of $\alpha$ defined in Eq.~(\ref{eq:betaF1}). Despite the large variations in the data, our model captures the average tendency of the empirical relation between $\beta$ and $\alpha$. We have verified that the statistical correlations between the values of $\beta$ and $\alpha$ are only statistically significant in the mid-range of $\alpha$ values, while there are no significant correlations within the lower and higher ranges of $\alpha$ values (Fig.~5 in S1 Appendix). Furthermore, the Akaike and Bayesian information criteria indicate that it is at least 100 times more likely that the empirical data come from our model (Eq.~\ref{eq:betaF1}) than from a null-model assuming no relationship between both exponents (intercept-only model: $\beta=\text{constant}$, see Fig.~4 in S1 Appendix). We also find that the estimated parameters from Eq.~(\ref{eq:betaF1}) are quite robust against thresholds for the total GDP; the adjusted values of $\gamma$, $\delta$, and $\delta$ barely change, even if we only consider the countries with top 50\% GDP values (Fig.~7 in S1 Appendix).

\begin{figure*}[!ht]
\begin{adjustwidth}{-2.25in}{0in}
\begin{center}
\includegraphics[width=1.4\textwidth,keepaspectratio]{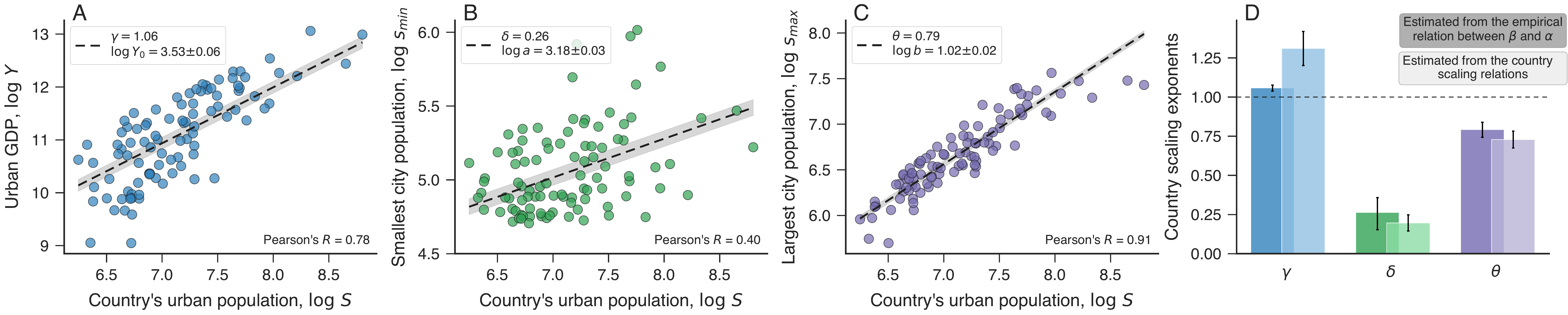}
\end{center}
\caption{{Country scaling relationships.} (A) Scaling relation between total urban GDP ($Y$) and total urban population ($S$) for all countries in our data set. (B) Scaling law between the smallest city population of a country ($s_{\text{min}}$) and the country's urban population ($S$). (C) Scaling law between the largest city population of a country ($s_{\text{max}}$) and the country's urban population ($S$). In these three panels, markers represent the values for each country and the dashed lines are the country scaling relationships of Eqs.~(\ref{eq:cntry_scaling}), (\ref{eq:cntry_scaling_smin}), and (\ref{eq:cntry_scaling_smax}), where the exponents $\gamma$, $\delta$, and $\theta$ are obtained from fitting the model of Eq.~(\ref{eq:betaF1}) to the empirical relation between $\beta$ and $\alpha$. Only the constants $Y_0$, $a$ and $b$ are adjusted to data and their best fitting values are shown in the panels ($\pm$ standard errors). (D) Comparison between estimates of the country scaling exponent obtained by fitting Eq.~(\ref{eq:betaF1}) to the empirical association between $\beta$ and $\alpha$ (bars in dark colors) and by directly fitting the values of the country scaling relationships (bars in light colors, see Fig.~6 in S1 Appendix for the adjusted scaling laws). Error bars represent 95\% bootstrap confidence intervals of the parameters. We notice that both approaches yield similar estimates, which are statistically indistinguishable in the cases of $\delta$ and $\theta$.}
\label{figure2}
\end{adjustwidth}
\end{figure*}

In addition to the previous model validation, the adjustment of Eq.~(\ref{eq:betaF1}) allows us to verify the consistency of our modeling approach through the country scaling relationships. Specifically, the country scaling exponents estimated from Eq.~(\ref{eq:betaF1}) should describe well the empirical country scaling relationships. To do so, we have adjusted the country scaling relationships of Eqs.~(\ref{eq:cntry_scaling_smin}), (\ref{eq:cntry_scaling_smax}), and (\ref{eq:cntry_scaling}) by considering only the prefactors ($a$, $b$, and $Y_0$) as fitting parameters and fixing country scaling exponents ($\delta$, $\theta$ and $\gamma$) to the values estimated from  Eq.~(\ref{eq:betaF1}). Figures~\ref{figure2}A-C show that the three country scaling relations describe quite well the empirical data. Furthermore, the country scaling exponents obtained by fitting Eq.~(\ref{eq:betaF1}) to the empirical relationship between $\beta$ and $\alpha$ agree well with the estimates directly obtained from the country scaling relations, that is, by fitting Eqs.~(\ref{eq:cntry_scaling_smin}), (\ref{eq:cntry_scaling_smax}), and (\ref{eq:cntry_scaling}) to the country-aggregated data, as shown in Fig.~\ref{figure2}D. The values of $\delta$ and $\theta$ estimated from both approaches are not significantly different, while the values of $\gamma$ are both above one but fitting the country scaling directly from data yields a slightly larger value. We believe this result provides support for our model, especially when considering the level of observed variation in the relationship between $\beta$ and $\alpha$ as well as that in the country scaling relations.

\subsection*{Simulating the association between both exponents}

The level of variation in the relation between the exponents $\beta$ and $\alpha$ (as well as in the country scaling relationships) is significantly high and hampers a more visual comparison with our model. In addition to indicating that our description of the association between both exponents is far from perfect, these variations also reflect possible estimation errors in the population and GDP values, errors related to the definition of city boundaries, and mainly errors associated with estimating the Zipf and urban scaling exponents (as data does not perfectly follow the Zipf's and urban scaling laws). As these errors emerge from different sources and do not seem to affect the association between $\beta$ and $\alpha$ systematically, we have treated them as random variations and investigated their role via an {\it in silico} experiment. As summarized in the Materials and Methods, we have generated artificial data at the city level (city population $s$ and city GDP $y$ values) by considering Zipf's law (Eq.~\ref{eq:zipf}), urban scaling (Eq.~\ref{eq:scaling}), country scaling relations (Eqs.~\ref{eq:cntry_scaling_smin} and \ref{eq:cntry_scaling_smax}), and our model (Eq.~\ref{eq:betaF1}). These simulations take as inputs the real values of total urban population $S$ and the exponent $\alpha$ of each country to generate replicas of these urban systems (set of values of $s$ and $y$), from which we can investigate the relationship between $\beta$ and $\alpha$ under different levels of variation in the urban and country scaling relations. To do so, we have fixed the variation intensity (that is, the standard deviation of Gaussian random variation around the scaling laws) in the simulated country scaling relations to a level comparable with the empirical data, and varied the variation intensity in the simulated urban scaling $\sigma_y$ in percentages of the value observed in the real data ($\sigma_y=100$\% means the standard deviation of the simulation is equal to the standard deviation in the empirical values of $\beta$). 

\begin{figure*}[!ht]
\begin{adjustwidth}{-2.25in}{0in}
\begin{center}
\includegraphics[width=1.4\textwidth,keepaspectratio]{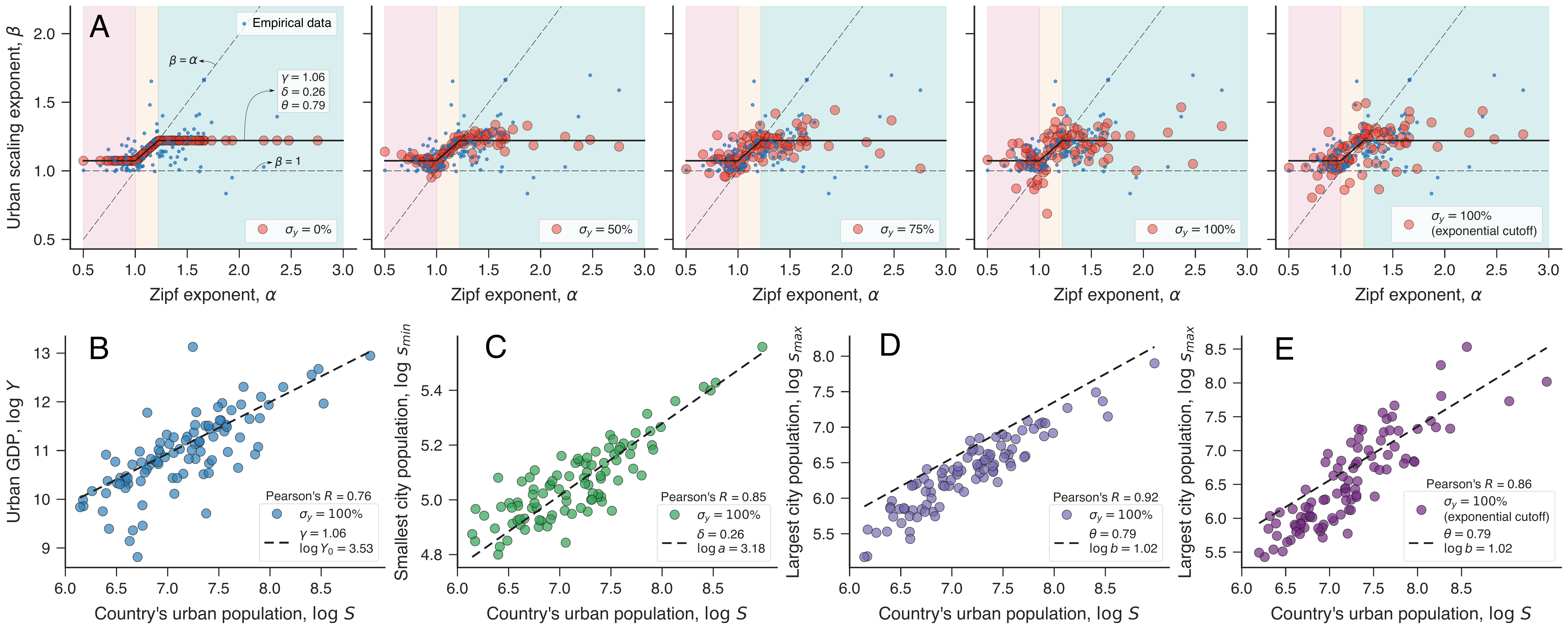}
\end{center}
\caption{{Simulating the connection between $\alpha$ and $\beta$, and the country scaling relationships.} (A) Simulated relationships between $\alpha$ and $\beta$ under different levels of random variation in the urban scaling law (Eq.~\ref{eq:scaling}). Here $\sigma_y$ is the percentage of the standard deviation in the empirical values of $\beta$. We have a perfect agreement between simulations and the model of Eq.~(\ref{eq:betaF1}) when $\sigma_y=0\%$, and the results become very similar to the empirical data (small blue dots) as the intensity of the random variation increases. (B)-(D) Simulated country scaling laws for $\sigma_y=100\%$. The dashed lines represent the scaling relationships of Eqs.~(\ref{eq:cntry_scaling}), (\ref{eq:cntry_scaling_smin}), and (\ref{eq:cntry_scaling_smax}), with parameters estimated from empirical data. We notice that the simulated scaling laws of total urban GDP ($Y$) and smallest city population ($s_\text{min}$) follow well the empirically adjusted relationships, while the simulated values for the largest city population ($s_\text{max}$) underestimate the empirical values. This occurs because large populations are rare and do not get close enough to the imposed maximum $s_\text{max}$. (E) Scaling relationship between $s_\text{max}$ and $S$ when considering that city population values are drawn from a power-law distribution with exponential cutoff. We notice this change makes the simulated results very similar to the empirical ones. The right-most plot in panel A shows a simulated relation between $\beta$ and $\alpha$ when considering the exponential cutoff.}
\label{figure3}
\end{adjustwidth}
\end{figure*}

The first four panels of Fig.~\ref{figure3}A show examples of simulated relationships (red circles) in comparison with empirical data (blue dots) for different values of $\sigma_y$. As expected, the simulated relationship perfectly agrees with our model (Eq.~\ref{eq:betaF1}) when there is no random variation in the urban scaling. More importantly, we note that the scattering of simulated data becomes visually very similar to the empirical data as the intensity of the random variation increases up to $\sigma_y=100$\% (Fig.~8 in S1 Appendix). We can also use the simulated data to corroborate our numerical experiment by verifying the country scaling relations. Figures~\ref{figure3}B-D show the simulated country scaling relations in comparison with the behavior of Eqs.~(\ref{eq:cntry_scaling_smin}), (\ref{eq:cntry_scaling_smax}), and (\ref{eq:cntry_scaling}) with parameters estimated from real data. We note that simulated scaling relations for total urban GDP ($Y$) and smallest city population ($\smin$) follow very closely the adjusted behavior of the empirical data. 

However, the simulated results for the largest city population ($\smax$) systematically underestimate the trend observed in the empirical data. This happens because, in our simulations, we have used a random number generator associated with a truncated power-law distribution (between $\smin$ and $\smax$) for mimicking population values according to Zipf's law. Since large populations are rare and the number of simulated cities is finite, the simulated values for the largest city population do not get close enough to the upper bound imposed by the truncated power-law distribution. Consequently, the simulated values of $\smax$ underestimate the empirical ones. We solve this issue by replacing the truncated power-law behavior by a power-law distribution with exponential cutoff (Materials and Methods). This modification does not alter the country scaling relations of total urban GDP and smallest city population, that is, the results of Fig.~\ref{figure3}B and \ref{figure3}C are not affected by the introduction of the exponential cutoff. Similarly, this modification does not affect the relationship between $\beta$ and $\alpha$ and the simulated associations with exponential cutoff are very similar to those obtained with the truncated power-law (as indicated by the right-most plot in Fig.~\ref{figure3}A). Indeed, the inclusion of this exponential cutoff only modifies the country scaling relation of the largest city population by increasing the simulated values of $\smax$. Figure~\ref{figure3}E shows an example of simulated results after after replacing the truncated cut-off with a exponential one for the country scaling relation between $\smax$ and $S$. We observe that the simulated values of $\smax$ obtained with the exponential cutoff are closer to the empirical data behavior than those obtained with the truncated power-law distribution (Fig.~\ref{figure3}D).

\section*{Discussion}
We have shown that the combination of Zipf's law and urban scaling implies a country scaling relationship, where the exponent is a function of the Zipf and the urban scaling exponents. While the Zipf and the urban scaling exponents vary from country to country, there is only one country scaling exponent for a given indicator, which in turn implies a direct association between the urban scaling and the Zipf exponents. In qualitative terms, our results agree with the more holistic idea that urban scaling exponents do not solely emerge from processes occurring within the city boundaries; instead, cities do not represent a closed and non-interacting system and what happens in the entire system (such as flow of people and goods) may affect urban scaling. Similarly to what happens in transportation theory where the product of two cities' populations is usually assumed to be proportional to the commuting flow between them (such as in gravity models), the population of cities is likely to represent an important factor for the interactions among cities. Under this assumption, the distribution of city population (summarized by the Zipf exponent) may thus represent an indirect proxy for interactions between cities, and the association observed between $\beta$ and $\alpha$ summarizes how the population distribution affects urban scaling and vice versa. 

Theoretically, the connection between the exponents $\beta$ and $\alpha$ would imply that instead of unrelated, Zipf's law and urban scaling are indeed the two sides of the same coin. However, the non-negligible variability observed in the empirical relationship do not corroborate with such a simple conclusion but suggest that other factors (such as level of socio-economic development and the particular history of an urban system) beyond population distribution may also have a significant effect on the urban scaling of GDP. Understanding the relative importance of population distribution on urban scaling (and vice versa) for different indicators is an important future contribution that is currently limited by the availability of other world-wide comparable city indicators. Even when considering the unexplained variation in the data, the connection between the two scaling laws uncovered by our work indicates the existence of universal processes governing both laws; however, finding out this commonality for arbitrary urban indicators still represents a challenging task.

In the context of urban GDP, our results show that urban systems with small values of the Zipf exponent also tend to present lower increasing returns to scale of GDP (low values of $\beta$). An urban system described by a small Zipf exponent has a more balanced population distribution, and consequently, fewer small cities and more large cities when compared with urban systems described by larger Zipf exponents. Thus, countries with proportionally more metropolises tend to have less pronounced increasing returns to scale than those having a small number of large cities. We hypothesize that urban systems with a large number of metropolises may also have a more integrated market whereby these large cities cooperate and develop specialized economic activities. As a result, urban systems with more metropolises would have a smaller degree of agglomeration of economic activities in large cities and so weaker increasing returns to scale for city GDP. In contrast, countries with a small number of large cities have to concentrate almost all complex economic activities in relatively fewer metropolises, which in turn intensifies the increasing returns to scale of urban GDP (high values of $\beta$). 

It is also interesting to note that high-GDP countries present relatively smaller values of Zipf and urban scaling exponents than mid-GDP and low-GDP countries (Fig.~\ref{figure1}A and Fig.~9 in S1 Appendix), that is, developed countries tend to have more metropolises and less pronounced increasing returns to scale of urban GDP. This latter point agrees with the more local observation that ``rich cities'' of the European Union (West cities) also exhibit smaller scaling exponents for GDP than their ``poor'' counterparts (East cities)~\cite{strano2016rich}. Large values of $\beta$ may express an urban system with high economic performance, but because $\beta$ alone does not define the total urban GDP, large values of $\beta$ also indicate a significant imbalance between the economic productivity of small and large cities. The association between $\beta$ and $\alpha$ also suggests that part of this economic inequality may reflect the unbalanced distribution of population. It is also worth mentioning the possibility of the existence of different urban planning regimes~\cite{hsieh2019housing} that may prevent sharp population agglomerations in developed countries and thus also partially explain the negative association between $\beta$ and total urban GDP. 

Our data do not allow a dynamic analysis nor the identification of the causal direction of the association between the exponents $\beta$ and $\alpha$. Still, a possible explanation for the observed differences among countries with different levels of development is that the economic policies in less developed countries have focused on large cities, fostering this unbalanced situation and creating cities larger than an hypothetical economically optimal. From a labor-market perspective, these large cities may attract inhabitants from smaller cities, changing both the urban scaling and the Zipf exponents. However, these new inhabitants may mostly find low-paying jobs or even become unemployed, which in turn might partially explain the poorer overall economic performance of less efficient urban systems. The association between both exponents is even more crucial because the world urban population may increase up to 90\% by 2100~\cite{jiangglobal2017}. This urbanization process is likely to be even more intense in developing countries and has the potential to further undermine their economic performance. Thus, it is not only important to discuss which scaling is desirable but also the population distribution within urban systems.

\section*{Materials and Methods}

\subsection*{Data}

The data used in this work is a product of two data sets. First, we use the recently released GHSL-FUA or eFUA~\cite{ghslfua2019schiavina,moreno2020metropolitan} definition of functional urban areas (FUAs). The eFUAs uses gridded population data from Global Human Settlement Layer (GHSL)~\cite{freire2015ghs,pesaresi2016global} and an automated classification approach for producing 9,031 urban boundaries (and population counts) over the entire globe (188 countries) for the year 2015. The eFUAs comprise high-density urban centers and their surrounding commuting zones and aim to capture the functional extent of cities. Second, we use the gridded GDP data provided by Kummu \textit{et al.}~\cite{kummu2018gridded}. This data set combines sub-national and national GDP data from different sources with population gridded data (from the GHSL and the HYDE 3.2~\cite{klein2010long}) to define three gridded global datasets: Gross Domestic Production per capita (5 arc-min resolution), Human Development Index (5 arc-min resolution), and Gross Domestic Production (30 arc-sec-min resolution). We have used only this latter file representing gridded values of total GDP with a resolution of 30 arc-sec (1~km at equator) for the year 2015 (the same information is also available for the years 1990 and 2000). To define GDP consistently at the grid level across countries, Kummu \textit{et al.} first calculate the GDP across gridcells within each subnational unit of a given country as the corresponding subnational GDP per capita value multiplied by the gridcell population. Next, the authors sum over these gridcell GDP values and divide by the sum gridcell populations in a country to define the population-weighted national GDP per capita. They then calculate the ratio between this population-weighted national GDP per capita and the subnational GDP per capita. Finally, they multiply this ratio by the national GDP per capita to obtain the final subnational GDP per capita values (that vary across subnational units in each country), which they then multiply by the population in each gridcell within each subnational unit to obtain the final GDP for every gridcell in a given country. This method ensures that the sum over GDP per capita values at the gridcell level always coincides with officially reported GDP per capita values for each country, and that there is global consistency because the method relies on secondary sources of reported subnational GDP per capita and internationally consistent population grids. The gridded GDP values are reported in 2011 international US dollars using purchasing power parity rates (total GDP-PPP). We overlay the gridded GDP data with eFUA boundary polygons and aggregate the GDP cell values within each polygon for associating a GDP value to each eFUA. We illustrate this procedure in Fig.~1 in S1 Appendix. Next, we group the resulting data by country and select the countries having at least 10 eFUAs. We also removed from our analysis 46 eFUA with null GDP (16 from India; 9 from Ethiopia; 5 from Pakistan; 3 from Sudan; 2 from Niger, Congo, and Chad; and 1 from Argentina, Benin, Egypt, Indonesia, Myanmar, Senegal, and Uganda). These criteria lead us to a data set comprising population $s$ and GDP $y$ of 8,650 functional urban areas from 96 countries.

\subsection*{Estimating the Zipf exponent}
Zipf's law (Eq.~\ref{eq:zipf}) implies that the complementary cumulative distribution function (CDF) of city population is a power law, $F(s)\sim s^{-\zeta}$, with $\zeta\approx\alpha$. We use this connection to estimate the values of $\alpha$ from the data. Specifically, we applied the approach of Clauset-Shalizi-Newman~\cite{clauset2009power} to obtain the exponent $\alpha$ via its maximum likelihood estimate
$\alpha = n / \left(\sum_{i=1}^n \ln {s_i}/{\tsmin}\right)\,,$ where $\tsmin$ is the lower bound of the power-law regime, $s_i$ is the population of the $i$-th city for a given country such that $s_i\geq\tsmin$, and $n$ is the number of city populations in the power-law regime. The value of $\tsmin$ is also estimated from data by minimizing the Kolmogorov-Smirnov ``distance'' between the empirical CDF of city populations and $F(s)$. The standard error in the Zipf exponent $\text{SE}_\alpha = \alpha/\sqrt{n}$ can be obtained from the width of the likelihood maximum~\cite{clauset2009power}. We have used the Clauset-Shalizi-Newman method as implemented in the Python module \texttt{powerlaw}~\cite{alstott2014powerlaw}. In addition to being a quite popular approach for fitting power-law distributions, Bhattacharya \textit{et al.}~\cite{bhattacharya2020consistency} have recently proven that the Clauset-Shalizi-Newman approach yields an unbiased and consistent estimator, that is, as data increases indefinitely the estimated parameters converge (in distribution) to the true values. We show the CDF and Zipf's law adjusted to each country in S1 Appendix, Sec.~4, where a good agreement is observed in the vast majority of cases. After estimating $\tsmin$, we filter out all cities with population smaller than this threshold in all other analyses, leading us to 4,571 functional urban areas from 96 countries. Thus, the urban scaling laws involve only cities belonging to the power-laws regime, and the countries' urban GDP ($Y$) and countries' urban population ($S$) are the aggregated values of urban GDP ($y_i$) and urban population ($s_i$) of cities belonging to the power-law regime for each country.

\subsection*{Estimating the urban scaling and the country scaling exponents}
Urban scaling and country scaling laws are generically represented by a power-law relation between a dependent variable $z$ and an independent variable $x$
\begin{equation}\label{eq:generic_PL}
    z = g\, x^\nu\,,
\end{equation}
where $g$ is a prefactor and $\nu$ is the power-law exponent. Equation~(\ref{eq:generic_PL}) is linearized via a logarithmic transformation
\begin{equation}\label{eq:generic_PL_log}
    \log z = \log g + \nu \log x\,,
\end{equation}
where $\log z$ and $\log x$ now represent the dependent and independent variables, $\log g$ is the intercept and $\nu$ the slope, both being regression coefficients of a corresponding linear model. We have estimated the values of $\log g$ and $\nu$ by adjusting Eq.~(\ref{eq:generic_PL_log}) to the log-transformed data via robust linear regression with the Huber loss function~\cite{huber2009robust}, as implemented in the Python module \texttt{statsmodels}~\cite{seabold2010statsmodels}. We further estimate standard errors and confidence intervals for the parameters $\log g$ and $\nu$ via bootstrapping~\cite{efron2009introduction}. We show the urban scaling law adjusted to each country in S1 Appendix, Sec.~4, where a good agreement is observed in the vast majority of cases. The adjusted country scaling laws are shown in Fig.~6 in S1 Appendix. In the case of Figs.~\ref{figure2}A-C, we have fixed the power-law exponents (regression coefficients) $\gamma$, $\delta$ and $\theta$ to values obtained from fitting Eq.~(\ref{eq:betaF1}) to the relation $\beta$ versus $\alpha$, and only the prefactors (intercepts of the linear model) $\log Y_0$, $\log a$, and $\log b$ have been considered as free parameters in the regression model.

\subsection*{Fitting our model to the relationship between $\beta$ and $\alpha$}
Our model for the relationship between $\beta$ and $\alpha$ is completely defined in S1 Appendix, Sec.~2. Depending on whether $\gamma>1$ or $\gamma<1$ and also on whether $\delta<\theta$ or $\delta>\theta$, we have an expression of $\beta$ as function of $\alpha$ (Eqs.~S56-S59, in S1 Appendix) and Eq.~(\ref{eq:betaF1}) is a particular case when $\gamma>1$ and $\delta<\theta$. We have adjusted the complete model (that is, without assuming anything about the parameters $\gamma$, $\delta$, and $\theta$) to the empirical relation between $\alpha$ and $\beta$ via the L-BFGS-B nonlinear optimization algorithm~\cite{byrd1995limited}, as implemented in the Python module \texttt{lmfit}~\cite{newvilleLMFIT2014} and without any constraint. The standard errors and the confidence intervals on the parameters $\gamma$, $\delta$ and $\theta$ are estimated via bootstrapping~\cite{efron2009introduction}. The best fitting parameters ($\pm$ standard errors) are $\gamma=1.06\pm0.01$, $\delta=0.26\pm0.05$, and $\theta=0.79\pm0.03$. This leads to Eq.~(\ref{eq:betaF1}) because the best fitting parameters yield $\gamma>1$ and $\delta<\theta$. Figures~10, and 11 in S1 Appendix depict different versions of Fig.~\ref{figure1}A where we label all countries and show the standard errors in $\beta$ and $\alpha$.

\subsection*{Simulating relations between $\beta$ and $\alpha$}
We simulate the relationship between $\beta$ and $\alpha$ by generating data at the city level. For a given country population $P$ with Zipf exponent $\alpha$, we start by generating a list of $m$ city populations $\mathcal{S} = \{s_1,\dots,s_i,\dots,s_m\}$ until satisfying the constraint $\sum_{i=1}^{m} s_i \approx S\,10^{\mathcal{N}(0,\sigma_\gamma)}$, where $\mathcal{N}(0,\sigma_\gamma)$ is a Gaussian random variable with zero mean and standard deviation $\sigma_\gamma$. Each $s_i$ is drawn from a power-law distribution $p(s)\sim s^{-(\alpha+1)}$ (compatible with Zipf's law) within the interval $\smin<s<\smax$, where $\smin$ and $\smax$ are obtained from the country scaling relations (Eqs.~\ref{eq:cntry_scaling_smin} and \ref{eq:cntry_scaling_smax}) with multiplicative random variations, that is, $\smin \sim S^\delta\,10^{\mathcal{N}(0,\sigma_\delta)}$ and $\smin \sim S^\theta\,10^{\mathcal{N}(0,\sigma_\theta)}$, where $\mathcal{N}(0,\sigma_\delta)$ and $\mathcal{N}(0,\sigma_\theta)$ are Gaussian random variables with zero mean and standard deviations $\sigma_\delta$ and $\sigma_\theta$, respectively. We next generate a list of urban indicators $\mathcal{Y} = \{y_1,\dots,y_i,\dots,y_m\}$, where $y_i = c\, s_i^\beta\, 10^{\mathcal{N}(0,\sigma_y)}$ and $\mathcal{N}(0,\sigma_y)$ is a Gaussian random variable with zero mean and standard deviation $\sigma_y$. In the expression for $y_i$, the value of $\beta$ is obtained from our model (Eq.~\ref{eq:betaF1}) while the value of $c$ is chosen to satisfy the condition $\sum_{i=1}^{m} y_i \approx Y$, where $Y$ is the total urban GDP (S1 Appendix, Sec.~3). The random variation controlled by the parameters $\sigma_\gamma$, $\sigma_\delta$, and $\sigma_\theta$ mimics the variability observed in the empirical country scaling relationships, and we set their values equal to the standard deviation of the bootstrap estimates of the country scaling exponents ($\gamma$, $\delta$, and $\theta$). On the other hand, the random variation controlled by $\sigma_y$ mimics the variability in the urban scaling relationships, and we set its value as a fraction of the standard deviation of the empirical values of $\beta$. We have thus applied this procedure by using all empirical values of $\alpha$ and $Y$ to obtain the simulated values of $\beta$, $Y$, $\smin$, $\smax$ from the lists $\mathcal{S}$ and $\mathcal{Y}$ under different values of $\sigma_y$ (Fig.~\ref{figure3}A). 

We have also considered a modification of this procedure where the population values were drawn from a power-law distribution with exponential cutoff~\cite{clauset2009power}, that is, $p(s)\sim s^{-(\alpha+1)}\exp(-s/s_0)$ ($s>\smin$), where $s_0$ is an additional parameter. This modification is necessary for reproducing the empirical behavior of the country scaling between $\smax$ and $S$. Because large city populations are very rare, the simulated values of $s_i$ obtained from the upper-truncated power-law distribution do not get close enough to the imposed maximum value ($\smax$). This results in the underestimation of $\smax$, as shown in Fig.~\ref{figure3}D. After replacing the upper-truncated behavior by the exponential cutoff, we note that the simulated country scaling of $\smax$ becomes very similar to the empirical relation (Fig.~\ref{figure3}E). In this simulation, we have chosen the value $s_0=3\times 10^7$ to make the simulated values of $\smax$ closer to the scaling law adjusted from the empirical data. It is worth mentioning that this change has no effect on the relationship between $\beta$ and $\alpha$ nor on the other country scaling relations (see the two right-most plots of Fig.~\ref{figure3}A). 

S1 Appendix, Sec.~3 shows more details on how we have implemented this simulation.

\section*{Acknowledgments}
We thank K.~Schmidheiny, J. Suedekum, and S.~Thies for useful discussions. This work emerged from ideas discussed at the symposium \emph{Cities as Complex Systems} (Hanover, July 13th-15th, 2016) which was generously funded by VolkswagenFoundation. HVR acknowledges the support of CNPq (Grants Nos. 440650/2014-3 and 303642/2014-9). DR thanks the Alexander von Humboldt Foundation for financial support under the Feodor Lynen Fellowship. DR is also grateful to the Leibniz Association (project IMPETUS) for financially supporting this research.

\section*{Supporting information}

\paragraph*{S1 Appendix.}
\label{S1_Appendix}
{\bf Supplementary Figures (1-11) and Supplementary Text (Sections 1-4) supporting the results discussed discussed in the main text.}

\bibliography{main_pone.bbl}

\clearpage
\setcounter{page}{1}
\setcounter{figure}{0}


\section*{Supplementary material}

S1 Appendix for

# Association between population distribution and urban GDP scaling

H. V. Ribeiro*, M. Oehlers, A. I. Moreno-Monroy, J. P. Kropp, D. Rybski*

*Corresponding authors. E-mail: hvr@dfi.uem.br or ca-dr@rybski.de

## Supplementary Figures

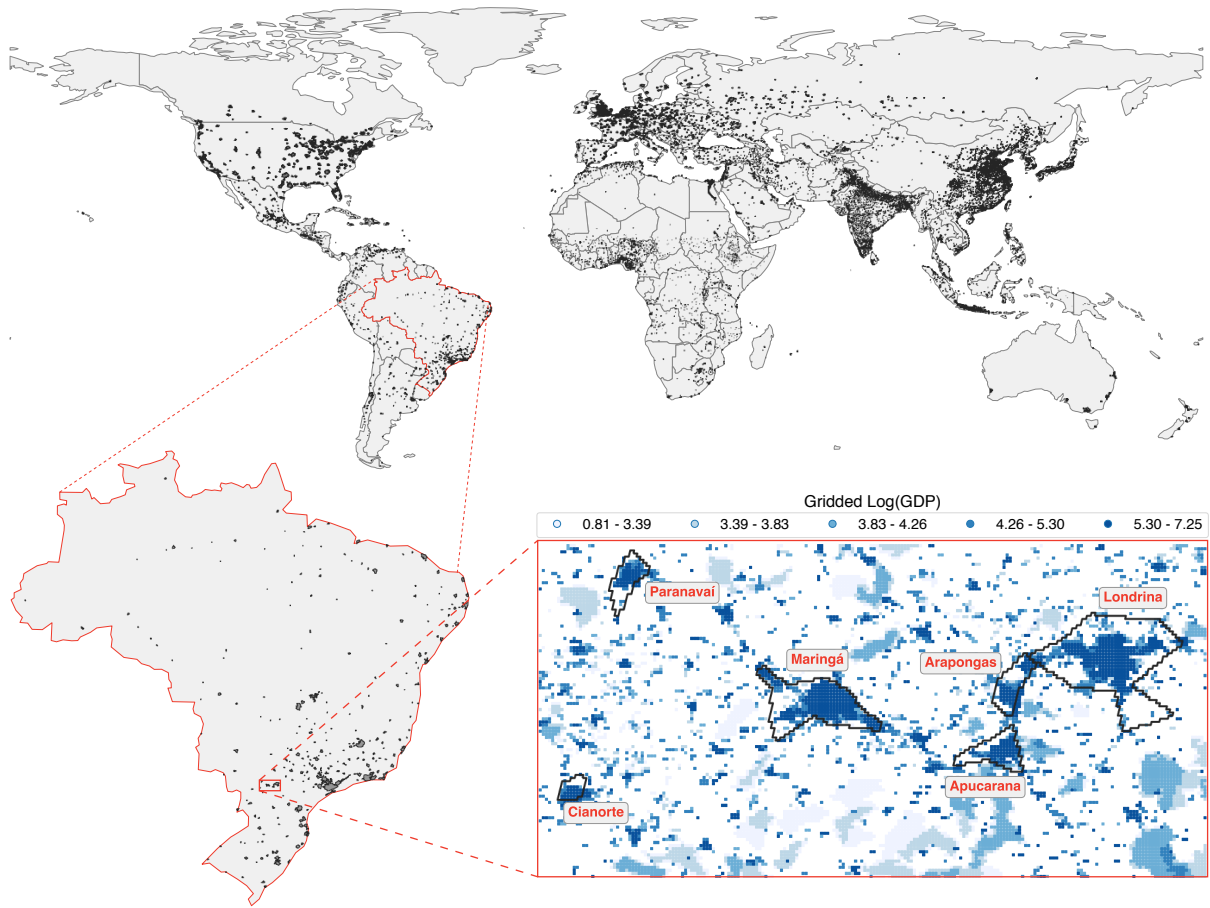

FIG. 1. **Illustration of the procedure used for obtaining population $s$ and GDP $y$ of cities over the entire globe.** The world map shows the boundaries of GHSL-OECD Functional Urban Areas (GHS-FUA or eFUA). This definition of urban areas has been recently released by the European Commission's science and knowledge service, with the aim of providing an internationally-comparable and harmonized definition of urban areas. The eFUA definition does not rely on local administrative boundary units and commuting flows data; instead, it uses gridded data from the Global Human Settlement Layer (GHSL) and an automated classification approach for producing urban boundaries [JRC Tech. Rep. (2019) and J. Urban. Econ., 103242 (2020)]. The shape file is provided for the year of 2015 together with population values for each eFUA. We combine eFUA boundary data with gridded GDP data provided by Kummu et al. [Sci. Data 5, 180004 (2018)]. In more detail, we use the total GDP-PPP (purchasing power parity, in constant 2011 international USD) data with 30 arc-sec resolution (≈1 km at the equator, file name: `GDP_PPP_30arcsec_v2.nc`) for the year 2015 to overlay with eFUA boundaries. Next, we aggregate all grid cell values within each eFUA boundary to associate a GDP value to each urban area of the world. We illustrate this idea by highlighting a small area in the southern part of Brazil, where we also plot the GDP grid cells.

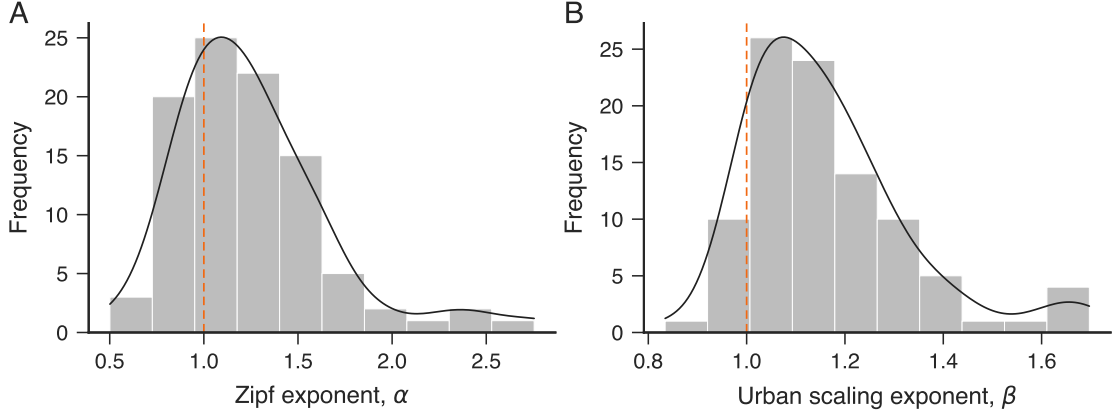

FIG. 2. **Probability distributions of the Zipf exponent $\alpha$ and the urban scaling exponent $\beta$.** (A) The bars represent a histogram for the values of $\alpha$ and the continuous line is a kernel density estimation (not normalized) for the same data. The vertical dashed line indicates the value $\alpha = 1$. The average value and the standard deviation of $\alpha$ are 1.24 and 0.38, respectively. (B) The bars represent a normalized histogram for the values of $\beta$ and the continuous line is a kernel density estimation (not normalized) for the same data. The vertical dashed line indicates the value $\beta = 1$. The average value and the standard deviation of $\beta$ are 1.16 and 0.17, respectively.

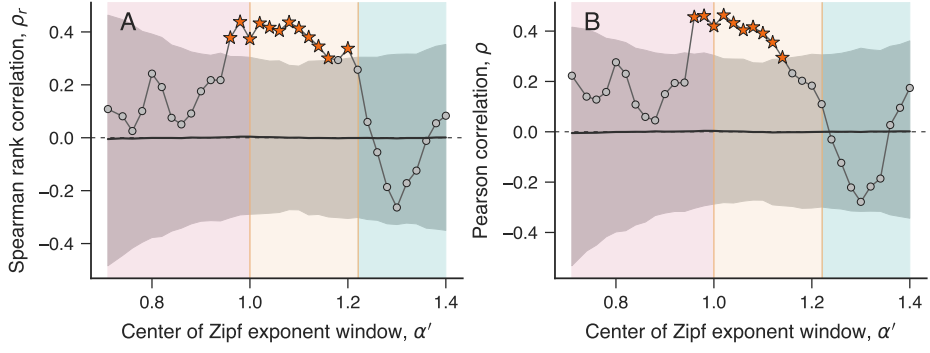

FIG. 3. **Correlation between the urban scaling exponent $\beta$ and the Zipf exponent $\alpha$ over sliding windows of values of $\alpha$.** (A) Spearman rank correlation ($\rho_r$) of the relationship between $\beta$ and $\alpha$ estimated within a sliding window centered in $\alpha'$. (B) Pearson correlation ($\rho$) of the relationship between $\beta$ and $\alpha$ estimated within a sliding window centered in $\alpha'$. For both correlation measures, the size of sliding windows is $\Delta\alpha' = 0.4$ and the step is 0.02. In addition, the red star markers indicate the correlation values that statistically significant at the 95% confidence level ($p$-values smaller than 0.05), while the gray circles represent non-significant correlation values ($p$-values larger than 0.05). Similar results are obtained for $\Delta\alpha'$ between 0.15 and 0.55. In both panels, the continuous black lines around zero represent average values of the correlation measures estimated after shuffling the values of $\beta$ (1000 realizations), and the gray shaded region stands for the 95% bootstrap confidence intervals. We notice that the correlation measures are larger and statistically significant for values $\alpha'$ around 1.1 and that the significant values cannot be explained by the distribution of $\beta$ values. This range containing the largest and significant values of $\rho$ and $\rho_r$ mostly correspond with mid-range of $\alpha$ values in Eq. (9) [Eq. S56 in Supplementary Text], where our model predicts a linear association between $\beta$ and $\alpha$ (see main text for details).

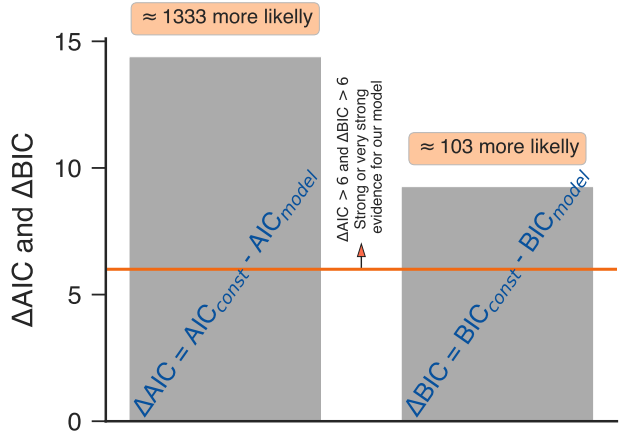

FIG. 4. **Comparison between the model of Eq. (9) [Eq. S56 in Supplementary Text] and a constant (intercept-only) model ($\beta$ = const).** The left bar shows the difference between the AIC (Akaike information criterion) values ($\Delta$AIC) estimated for an intercept-only model (AIC$_{\text{const}}$) and for the model of Eq. (9) [Eq. S56 in Supplementary Text] (AIC$_{\text{model}}$). The model with the lowest AIC value gives a better description of the data. Thus, $\Delta$AIC > 0 indicates that the model of Eq. (9) [Eq. S56 in Supplementary Text] is a better description for data. As a rule of thumb, it is common to consider 6 < $\Delta$AIC < 10 as a strong evidence and $\Delta$AIC $\geq$ 10 as a very strong evidence (our case, since $\Delta$AIC $\approx$ 14.4) in favor of the model of Eq. (9) [Eq. S56 in Supplementary Text]. In addition, because the AIC value is proportional to the negative of log-maximum-likelihood of the model, the quantity $e^{\Delta\text{AIC}}$ approximates how more likely the model of Eq. (9) [Eq. S56 in Supplementary Text] is to describe the data when compared with an intercept-only model. Thus, we notice that Eq. (S56) is $\approx$ 1333 times more likely to describe the relationship between $\beta$ and $\alpha$ than the intercept-only model. The right bar shows the difference between the BIC (Bayesian information criterion) values ($\Delta$BIC) estimated for an intercept-only model (BIC$_{\text{const}}$) and for the model of Eq. (9) [Eq. S56 in Supplementary Text] (BIC$_{\text{model}}$). The interpretation of BIC and AIC values are analogous. The results indicate a strong evidence ($\Delta$BIC $\approx$ 9.3) in favor of Eq. (9) [Eq. S56 in Supplementary Text] and our model is $\approx$ 103 times more likely to describe the empirical data than the intercept-only model. It is worth remembering that both AIC and BIC values account for the fact that Eq. (9) [Eq. S56 in Supplementary Text] has more free parameters than the intercept-only model (3 versus 1 parameter); moreover, BIC values penalize the number of parameters more heavily.

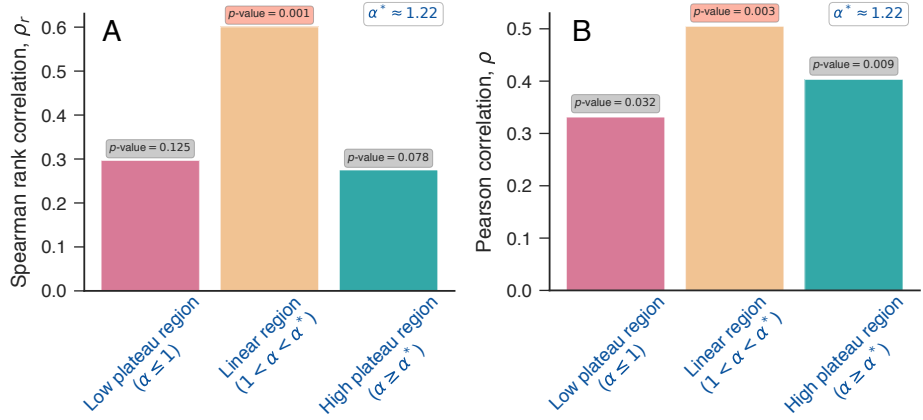

FIG. 5. **Correlations in each range of values of $\alpha$ defined in Eq. (9) [Eq. S56 in Supplementary Text]**. (A) Spearman rank correlation $\rho_r$ between $\beta$ and $\alpha$ in the low plateau ($\alpha \leq 1$), linear regime ($1 < \alpha < \alpha^*$), and high plateau ($\alpha \geq \alpha^*$) regions of Fig. 1 in the main text. Panel (B) shows the same for the Pearson correlation coefficient $\rho$. In both panels, we indicate above the bars the $p$-values testing the significance of the correlation coefficient (permutation test). We notice that $\rho_r$ and $\rho$ are significantly higher in the linear region. We further observe that the correlations are statistically significant at 99% confidence level only in the linear region. Here $\alpha^* = 1 + \frac{\gamma-1}{\delta} \approx 1.22$ [Eq. (9) of the main text or Eq. S56 in Supplementary Text, and Fig. 15].

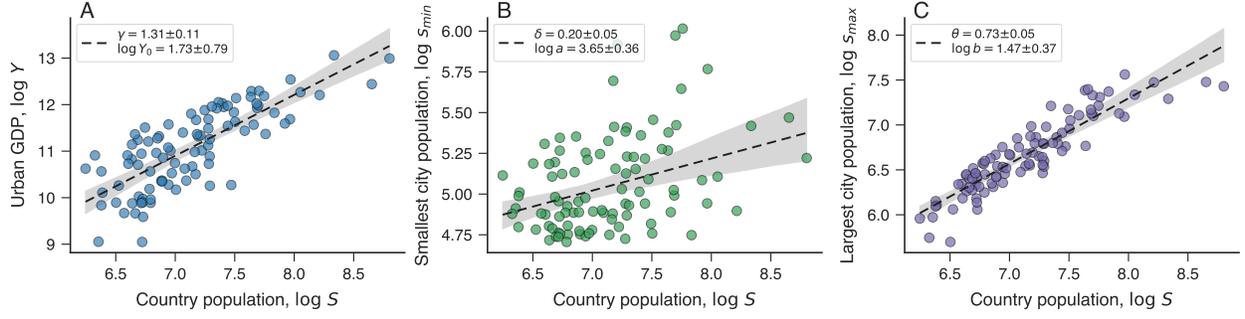

FIG. 6. **Fitting country scaling relationships directly to data.** (A) Scaling relation between total urban GDP of countries ($Y$) and country's urban population values ($S$). (B) Scaling law between the smallest city population of a country ($s_{\min}$) and country's urban population ($S$). (C) Scaling law between the largest city population of a country ($s_{\max}$) and country's urban population ($S$). In the three panels markers represent the values for each country and the dashed lines are the country scaling relationships of Eqs. (8), (5), and (6) [Eqs. (S6), (S7), and (S8) in Supplementary Text], where the exponents $\gamma$, $\delta$, and $\theta$ (and the prefactors $\log Y_0$, $\log a$, and $\log b$) are obtained via robust linear regression of the log-transformed data. The best fitting parameters are shown in the panels (± standard errors) and gray shaded regions stand for the 95% confidence band of the models. As discussed in the main text, the exponents estimated directly from the scaling laws are very similar to the values obtained by fitting Eq. (9) [Eq. S56 in Supplementary Text] to the relationship between $\beta$ and $\alpha$ (see Fig. 2D of main text for a comparison).

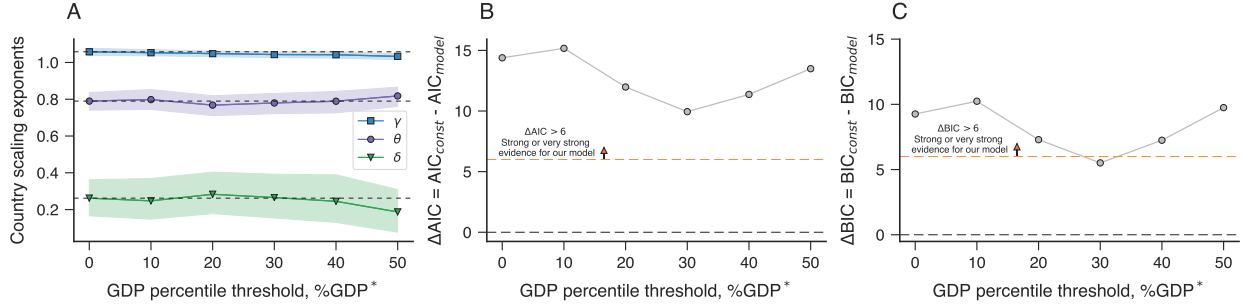

FIG. 7. **Robustness of the model of Eq. (9) [Eq. S56 in Supplementary Text] against different GDP threshold.** (A) The colored markers show the values of $\gamma$, $\delta$, and $\theta$ estimated by fitting (via the L-BFGS algorithm) Eq. (9) [Eq. S56 in Supplementary Text] to the relationship between $\beta$ and $\alpha$ after selecting countries with total urban GDP higher than the percentile GDP threshold (%GDP$^*$). The shaded regions are 95% bootstratp confidence intervals and the dashed lines represent the values when considering all data. We notice the exponent estimates are quite stable. (B) Difference between the AIC (Akaike information criterion) values ($\Delta$AIC) estimated for an intercept-only model (AIC$_{\text{const}}$) and for the model of Eq. (9) [Eq. S56 in Supplementary Text] (AIC$_{\text{model}}$) as a function of the percentile GDP threshold (%GDP$^*$). (C) Difference between the BIC (Bayesian information criterion) values ($\Delta$BIC) estimated for an intercept-only model (BIC$_{\text{const}}$) and for the model of Eq. (9) [Eq. S56 in Supplementary Text] (BIC$_{\text{model}}$) as a function of the percentile GDP threshold (%GDP$^*$). Results show strong evidence ($\Delta$AIC > 6 and $\Delta$BIC > 6) for the model of Eq. (9) [Eq. S56 in Supplementary Text] at practically all threshold values.

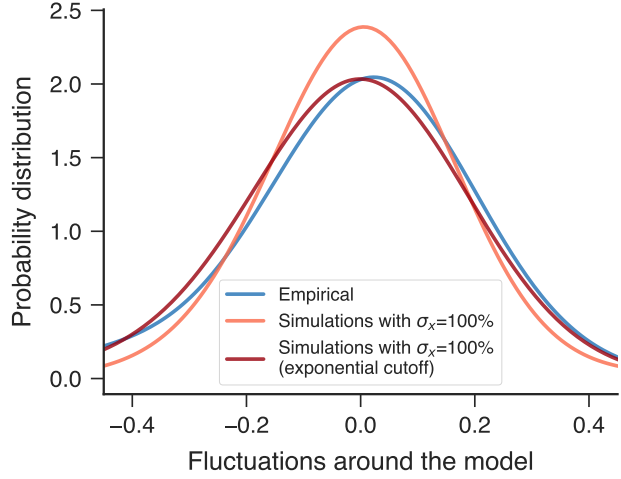

FIG. 8. **Probability distributions of the random variations around the model of Eq. (9) [Eq. S56 in Supplementary Text].** The different curves show kernel density estimations of the random variations around the model of Eq. (9) when considering the empirical data and the simulations obtained for $\sigma_y = 100\%$. We observe that the distributions obtained from the simulations are very similar to the one emerging from the empirical data, especially when including an exponential cutoff in the power-law distribution associated with the simulated city sizes.

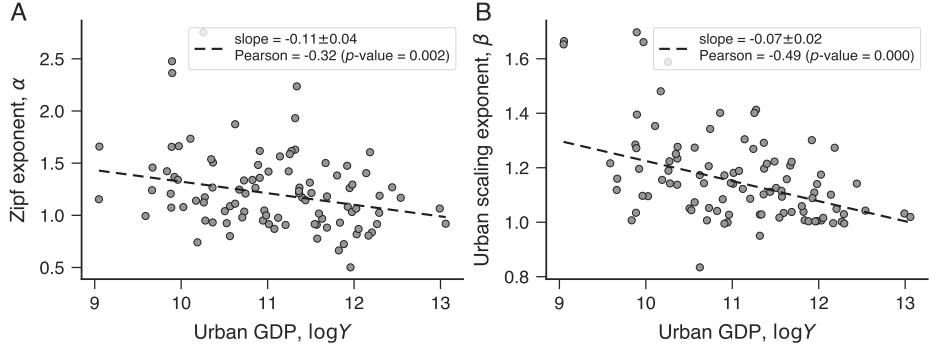

FIG. 9. **Association between the exponents $\alpha$ and $\beta$ and total urban GDP $Y$.** (A) Scatter plot of Zipf exponent $\alpha$ versus the total urban GDP $Y$. (B) Scatter plot of urban scaling exponent $\beta$ versus the total urban GDP $Y$. The dashed line in each panel represents a linear regression with slope indicated within the plots (where we also shoe the Pearson correlation coefficient and the corresponding $p$-values). We observe that both exponents also significantly associated with urban GDP $Y$, such that an increase in $Y$ tends to be followed by a decrease in $\alpha$ and $\beta$.

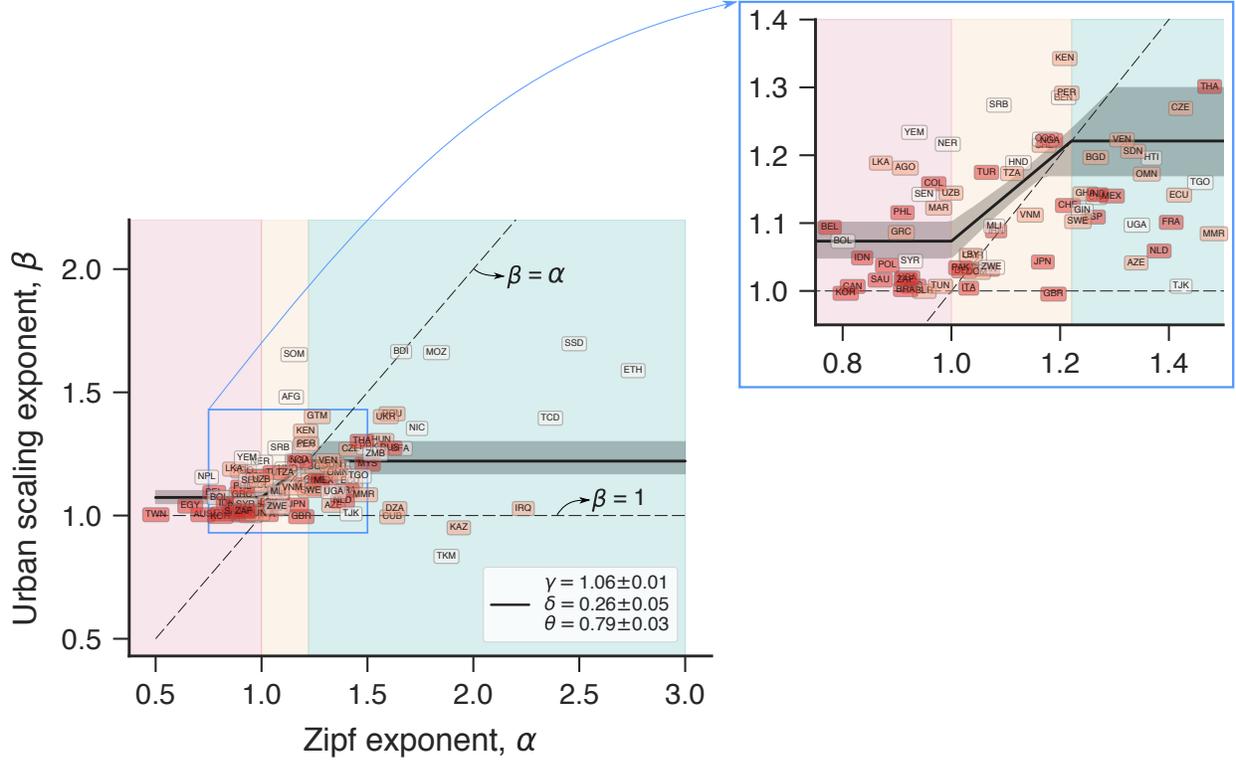

FIG. 10. **Association between the urban scaling exponent $\beta$ and the Zipf exponent $\alpha$.** The markers indicate the values of $\beta$ versus $\alpha$ for each country in our data set. The three marker colors group countries according to the tercile values of the total urban GDP distribution (for instance, High-GDP countries have highest ≈ 33% GDP values, as the main text). The horizontal dashed line shows the $\beta = 1$, while the inclined dashed line represents the $\beta = \alpha$ relationship. The continuous line shows the model of Eq. (9) [Eq. S56 in Supplementary Text] adjusted to the data and the gray shaded region stands for the 95% bootstrap confidence band. The colored shaded regions represent the different intervals of $\alpha$ defined in Eq. (9) [Eq. S56 in Supplementary Text]. All countries are labeled according to their ISO codes (the three letters over each marker). The blue frame indicates a region that has been magnified on the right.

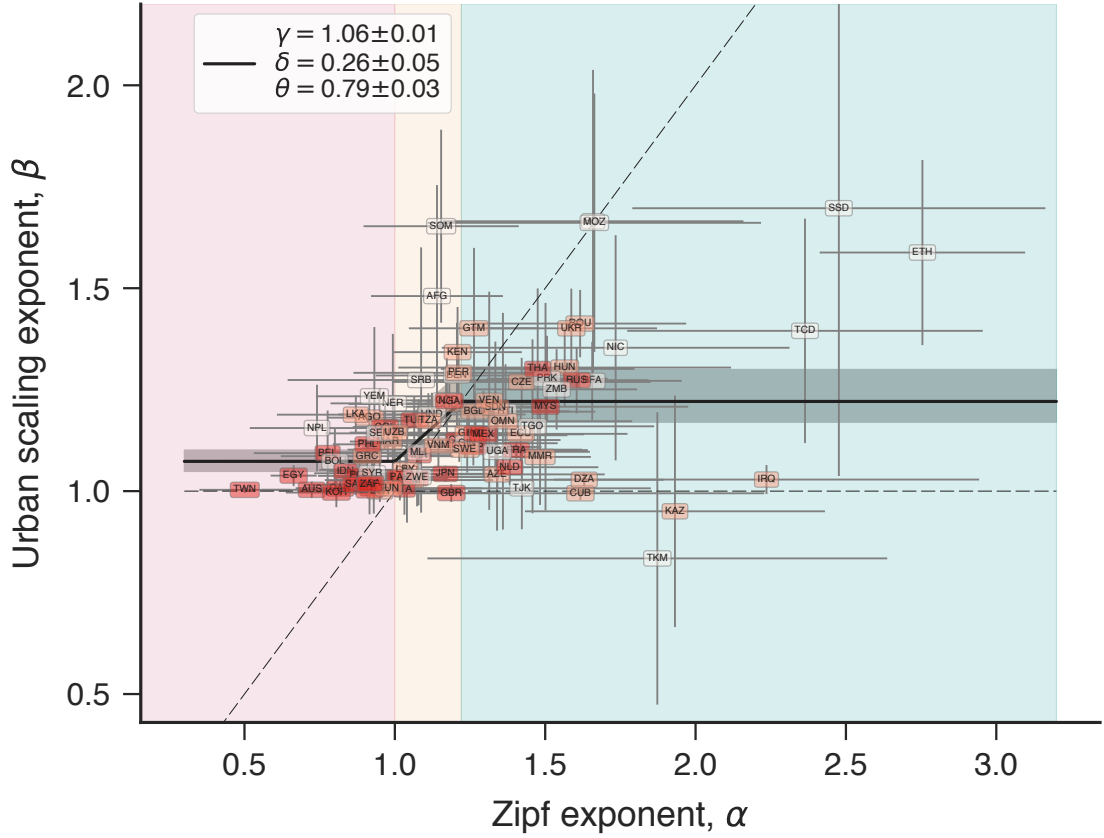

FIG. 11. **Association between the urban scaling exponent $\beta$ and the Zipf exponent $\alpha$.** The same as Fig. (10) but with error bars indicating standard errors. The three colors group countries according to the tercile values of the total urban GDP distribution (for instance, High-GDP countries have highest ≈ 33% GDP values). The horizontal dashed line shows the $\beta = 1$, while the inclined dashed line represents the $\beta = \alpha$ relationship. The continuous line shows the model of Eq. (9) [Eq. S56 in Supplementary Text] adjusted to data and the gray shaded region stands for the 95% bootstrap confidence band. The colored shaded regions represent the different intervals of $\alpha$ defined in Eq. (9) [Eq. S56 in Supplementary Text]. All countries are labeled according to their ISO codes (the three letters over each marker).

# Supplementary Text

## 1. CONNECTION BETWEEN URBAN SCALING AND ZIPF EXPONENTS

Let us start by considering an urban system (a country) composed of $m$ cities with populations $\{s_1, s_2, \ldots, s_m\}$. Suppose now that we sort these population values $s$ in descending order $\{s_{\max}, \ldots, s, \ldots, s_{\min}\}$ (where $s_{\max}$ and $s_{\min}$ are respectively the largest and smallest city population in the system) and associate a rank variable $r$ defined by $\{1, 2, \ldots, r, \ldots, m\}$ to each city population. Zipf's law for cities states that rank $r$ and city population $s$ are related via

$$r \sim s^{-\alpha}, \tag{S1}$$

where $\alpha > 0$ is a power-law exponent or the Zipf exponent. Zipf's law implies that the complementary cumulative distribution of city population $F(s)$ can be written as $F(s) \sim s^{-\zeta}$, with $\zeta \approx \alpha$, and, consequently, the probability density distribution $P(S)$ of city population is

$$P(s) \sim s^{-(\alpha+1)} \quad \text{for } s_{\min} < s < s_{\max}. \tag{S2}$$

From Eq. (S2) we can also write the frequency function or the absolute number of cities with population $s$ as

$$p(s) = ks^{-(\alpha+1)} \quad \text{for } s_{\min} < s < s_{\max}, \tag{S3}$$

where $k$ is a normalization constant determined by

$$S = \sum_{s=s_{\min}}^{s_{\max}} sp(s) \approx k \int_{s_{\min}}^{s_{\max}} s^{-\alpha} ds \tag{S4}$$

and $S$ is the total population of the urban system.

On the other hand, the urban scaling hypothesis establishes a power-law relationship between an urban indicator $y$ (*e.g.* the urban GDP) and the population $s$ of a city, that is,

$$y = cs^{\beta}, \tag{S5}$$

where $c$ is a positive constant, and $\beta > 0$ is the urban scaling exponent. Our work has shown that the values of $\beta$ and $\alpha$ are related to each other, and

we propose that the relationship between these two exponents emerge from country scaling relationships.

To derive the relationship between $\beta$ and $\alpha$, we consider $S$ as the total population of an urban system and $Y$ as the total quantity of the city indicator $y$ for the whole urban system (*e.g.* total urban GDP). It is worth noticing that the values of $S$ and $Y$ represent the empirical values of total population and total GDP of all cities of a country within the power-law regime (that is, all cities with population larger than the lower cutoff obtained after fitting the Zipf's law to the data, see Material and Methods in the main text for details). In addition to the urban scaling that considers relations between population and indicators of cities within one country, we further assume three different country scaling relations between population and indicators across countries. The first one is a generalization of the urban scaling hypothesis at the country level

$$Y = Y_0 S^\gamma, \qquad (S6)$$

where $Y_0$ is a constant, and $\gamma$ is the country scaling exponent associated with the indicator $Y$. The other two relationships relate the population of the smallest ($s_{\min}$) and largest ($s_{\max}$) cities in an urban system to the total population S, that is,

$$s_{\min} = aS^\delta \qquad (S7)$$
$$s_{\max} = bS^\theta, \qquad (S8)$$

where $a$ and $b$ are positive constants, and $\delta > 0$ and $\theta > 0$ are two additional country scaling exponents. We have empirically verified these three country scaling relationships in our data (Fig. 6 and Fig. 2 of the main text); however and as we shall prove, the relationship of Eq. (S6) can also be derived from Zipf's law (Eq. S3), urban scaling (Eq. S5) and the two country scalings for $s_{\min}$ and $s_{\max}$ (Eqs. S7 and S8). Qualitatively speaking, we consider that the existence of country scaling relationships constrains the population and urban indicator values in a country which in turn ties the values of $\beta$ and $\alpha$.

To find this connection, we notice that the total indicator $Y$ can be ex-

pressed by using Zipf's law and urban scaling, that is,

$$Y = \sum_{s=s_{\min}}^{s_{\max}} yp(s) = \sum_{s=s_{\min}}^{s_{\max}} cs^\beta p(s) \approx kc \int_{s_{\min}}^{s_{\max}} s^{\beta-\alpha-1} ds. \tag{S9}$$

By determining the constant $k$ from Eq. (S4) and solving the integral in Eq. (S9), we should obtain the country scaling of Eq. (S6) in terms of the exponents $\alpha$ and $\beta$. Thus, if we assume $\gamma$, $\delta$, and $\theta$ to be fixed, the changes in $\alpha$ and $\beta$ must be constrained in order to yield the particular value of $\gamma$ imposed by country scaling in Eq. (S6).

**Case A:** $\alpha \neq \beta \neq 1$

To start our calculations, we first consider the general case where $\alpha \neq \beta \neq 1$. Under this assumption, Eq. (S4) yields

$$k = \frac{(\alpha-1) S s_{\max}^\alpha s_{\min}^\alpha}{s_{\max}^\alpha s_{\min} - s_{\max} s_{\min}^\alpha}, \tag{S10}$$

and then by replacing $s_{\min}$ with Eq. (S7) and $s_{\max}$ with Eq. (S8), we find

$$k = \frac{(\alpha-1)(ab)^\alpha S S^{(\delta+\theta)\alpha}}{ab^\alpha S^{\delta+\theta\alpha} - a^\alpha b S^{\theta+\delta\alpha}}. \tag{S11}$$

By solving the integral in Eq. (S9), we find

$$Y = \frac{kc}{\beta-\alpha} \left( s_{\max}^{\beta-\alpha} - s_{\min}^{\beta-\alpha} \right), \tag{S12}$$

and after replacing $k$ by Eq. (S11), $s_{\min}$ by Eq. (S7), and $s_{\max}$ by Eq. (S8), we have

$$\begin{aligned} Y &= \frac{c(\alpha-1)S}{\alpha-\beta} \left( \frac{a^\beta b^\alpha S^{\delta\beta+\theta\alpha} - a^\alpha b^\beta S^{\theta\beta+\delta\alpha}}{ab^\alpha S^{\delta+\theta\alpha} - a^\alpha b S^{\theta+\delta\alpha}} \right) \\ &= \frac{c(\alpha-1)S}{\alpha-\beta} \left( \frac{a^\beta b^\alpha S^{\lambda_1} - a^\alpha b^\beta S^{\lambda_2}}{ab^\alpha S^{\lambda_3} - a^\alpha b S^{\lambda_4}} \right) \\ &= \frac{c(\alpha-1)S}{\alpha-\beta} \left( \frac{\Lambda_1 - \Lambda_2}{\Lambda_3 - \Lambda_4} \right), \end{aligned} \tag{S13}$$

where

$$\begin{aligned} \lambda_1 &= \delta\beta + \theta\alpha \\ \lambda_2 &= \theta\beta + \delta\alpha \\ \lambda_3 &= \delta + \theta\alpha \\ \lambda_4 &= \theta + \delta\alpha. \end{aligned} \tag{S14}$$

15

and

$$\begin{aligned}\Lambda_1 &= a^\beta b^\alpha S^{\lambda_1}\\ \Lambda_2 &= a^\alpha b^\beta S^{\lambda_2}\\ \Lambda_3 &= ab^\alpha S^{\lambda_3}\\ \Lambda_4 &= a^\alpha b S^{\lambda_4}\,.\end{aligned} \qquad (S15)$$

Because $S$ represents the total population of an urban system or a country, we take the limit of large $S$ in Eq. (S13). To do so, we need to know the conditions under which $\Lambda_1$ dominates over $\Lambda_2$ in the numerator of Eq. (S13), that is, when

$$\begin{aligned}\lambda_1 &> \lambda_2\\ \delta\beta + \theta\alpha &> \theta\beta + \delta\alpha\\ (\delta - \theta)(\beta - \alpha) &> 0\,,\end{aligned} \qquad (S16)$$

and also the conditions under which $\Lambda_3$ dominates over $\Lambda_4$ in the denominator of Eq. (S13), that is, when

$$\begin{aligned}\lambda_3 &> \lambda_4\\ \delta + \theta\alpha &> \theta + \delta\alpha\\ (\delta - \theta)(1 - \alpha) &> 0\,.\end{aligned} \qquad (S17)$$

By investigating the inequality in Eq. (S16), we end up with the decision tree of Figure 12. Similarly, we find the decision tree of Figure 13 for the inequality in Eq. (S17).

Having the results of Figures 12 and 13, we can find the behavior of $Y$ for large $S$ in Eq. (S13) for each of the following conditions.

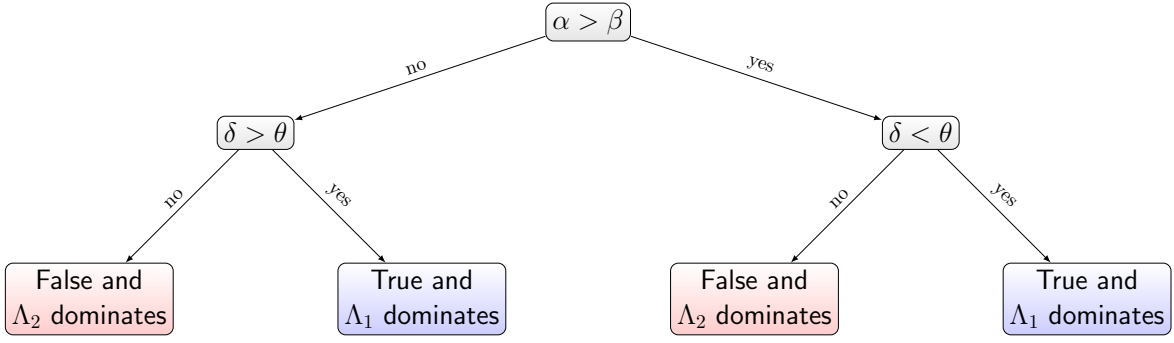

FIG. 12. Decision tree associated with the inequality $\lambda_1 > \lambda_2$ (Eq. S16) that defines whether $\Lambda_1$ dominates over $\Lambda_2$ (or vice-versa) in Eq. (S13).

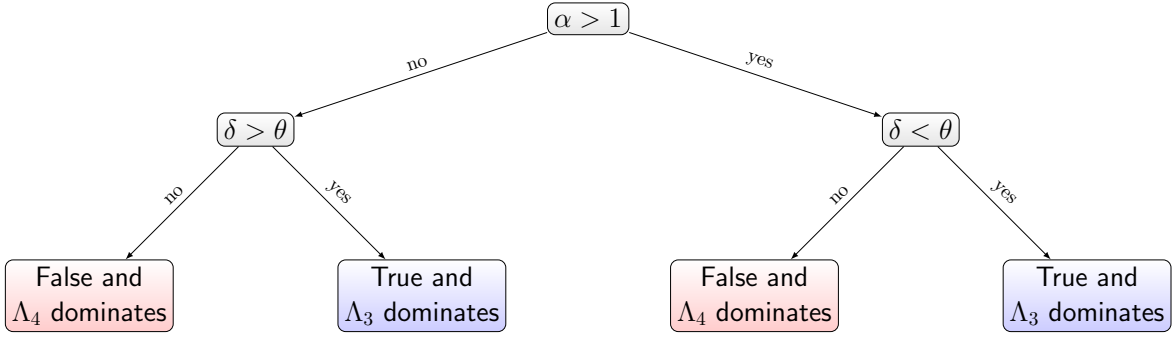

FIG. 13. Decision tree associated with the inequality $\lambda_3 > \lambda_4$ (Eq. S17) that defines whether $\Lambda_3$ dominates over $\Lambda_4$ (or vice-versa) in Eq. (S13).

**Case A.1:** $\alpha > \beta$, $\alpha > 1$, and $\delta < \theta$

Under these conditions, we find that $\Lambda_1$ and $\Lambda_3$ dominate in Eq. (S13), yielding

$$Y \approx \frac{c(\alpha-1)a^{\beta-1}}{\alpha-\beta}S^{1+\lambda_1-\lambda_3}$$
$$\approx \frac{c(\alpha-1)a^{\beta-1}}{\alpha-\beta}S^{1-\delta+\beta\delta}. \tag{S18}$$

By comparing the result of Eq. (S18) with the country scaling in Eq. (S6), we find

$$\beta = 1 + \frac{\gamma-1}{\delta} \qquad \text{for } \{\alpha > \beta \ \& \ \alpha > 1 \ \& \ \delta < \theta\}. \tag{S19}$$

It is worth noticing that the previous conditions depend on $\beta$, but by com-

17

bining $\alpha > \beta$ & $\alpha > 1$ and eliminating $\beta$, we can rewrite Eq. (S19) as

$$\beta = 1 + \frac{\gamma - 1}{\delta} \qquad \text{for } \left\{\alpha > \max\left(1, 1 + \frac{\gamma - 1}{\delta}\right) \ \& \ \delta < \theta\right\}. \tag{S20}$$

Equation (S20) thus predicts that $\beta$ is a constant, that is, not dependent on $\alpha$ when $\{\alpha > \max\left(1, 1 + \frac{\gamma-1}{\delta}\right) \ \& \ \delta < \theta\}$.

**Case A.2:** $\alpha > \beta$, $\alpha < 1$, and $\delta < \theta$

For this case, we find that $\Lambda_1$ and $\Lambda_4$ dominate the behavior of Eq. (S13) for large $S$, yielding

$$\begin{aligned} Y &\approx \frac{c(1-\alpha)a^{\beta-\alpha}b^{\alpha-1}}{\alpha - \beta} S^{1+\lambda_1-\lambda_4} \\ &\approx \frac{c(1-\alpha)a^{\beta-\alpha}b^{\alpha-1}}{\alpha - \beta} S^{1-\theta+\beta\delta+\alpha(\theta-\delta)}. \end{aligned} \tag{S21}$$

By comparing the previous result with Eq. (S6), we find

$$\beta = \frac{\gamma + \theta - 1}{\delta} + \left(1 - \frac{\theta}{\delta}\right)\alpha \qquad \text{for } \{\alpha > \beta \ \& \ \alpha < 1 \ \& \ \delta < \theta\}, \tag{S22}$$

and after eliminating $\beta$ from the condition, we have

$$\beta = \frac{\gamma + \theta - 1}{\delta} + \left(1 - \frac{\theta}{\delta}\right)\alpha \qquad \text{for } \left\{\gamma < 1 \ \& \ 1 + \frac{\gamma - 1}{\theta} < \alpha < 1 \ \& \ \delta < \theta\right\}. \tag{S23}$$

Equation (S23) thus predicts a linear decreasing correspondence between $\beta$ and $\alpha$ when $\{\gamma < 1 \ \& \ 1 + \frac{\gamma-1}{\theta} < \alpha < 1 \ \& \ \delta < \theta\}$.

**Case A.3:** $\alpha < \beta$, $\alpha > 1$, and $\delta < \theta$

Under these assumptions, $\Lambda_2$ and $\Lambda_3$ dominate the behavior of Eq. (S13) for large $S$, leading to

$$\begin{aligned} Y &\approx \frac{c(\alpha-1)a^{\alpha-1}b^{\beta-\alpha}}{\beta - \alpha} S^{1+\lambda_2-\lambda_3} \\ &\approx \frac{c(\alpha-1)a^{\alpha-1}b^{\beta-\alpha}}{\beta - \alpha} S^{1-\delta+\beta\theta+\alpha(\delta-\theta)}. \end{aligned} \tag{S24}$$

By comparing the previous result with Eq. (S6), we find

$$\beta = \frac{\gamma + \delta - 1}{\theta} + \left(1 - \frac{\delta}{\theta}\right)\alpha \qquad \text{for } \{\alpha < \beta \ \& \ \alpha > 1 \ \& \ \delta < \theta\}, \tag{S25}$$

that after eliminating $\beta$ from the condition leads to

$$\beta = \frac{\gamma + \delta - 1}{\theta} + \left(1 - \frac{\delta}{\theta}\right)\alpha \qquad \text{for } \left\{\gamma > 1 \ \& \ 1 < \alpha < 1 + \frac{\gamma - 1}{\delta} \ \& \ \delta < \theta\right\}. \tag{S26}$$

Similarly to Eq. (S23), the result of Eq. (S26) predicts a linear increasing correspondence between $\beta$ and $\alpha$ when $\{\gamma > 1 \ \& \ 1 < \alpha < 1 + \frac{\gamma-1}{\delta} \ \& \ \delta < \theta\}$. It is worth noticing that the conditions underlying Eqs. (S23) and (S26) are mutually exclusive, that is, only one of the linear correspondences exists depending on whether $\gamma > 1$ (super-linear country scaling, Eq. S26) or $\gamma < 1$ (sub-linear country scaling, Eq. S23).

**Case A.4:** $\alpha < \beta$, $\alpha < 1$, and $\delta < \theta$

For this case, we find that $\Lambda_2$ and $\Lambda_4$ dominate the behavior of Eq. (S13) for large $S$, resulting in

$$\begin{aligned} Y &\approx \frac{c(1-\alpha)b^{\beta-1}}{\beta - \alpha} S^{1+\lambda_2-\lambda_4} \\ &\approx \frac{c(1-\alpha)b^{\beta-1}}{\beta - \alpha} S^{1-\theta+\beta\theta}. \end{aligned} \tag{S27}$$

By comparing the previous result with Eq. (S6), we find

$$\beta = 1 + \frac{\gamma - 1}{\theta} \qquad \text{for } \{\alpha < \beta \ \& \ \alpha < 1 \ \& \ \delta < \theta\}, \tag{S28}$$

and after eliminating $\beta$ from the condition, we have

$$\beta = 1 + \frac{\gamma - 1}{\theta} \qquad \text{for } \left\{\alpha < \min\left(1, 1 + \frac{\gamma - 1}{\theta}\right) \ \& \ \delta < \theta\right\}. \tag{S29}$$

Equation (S29) indicates that $\beta$ is a constant when $\{\alpha < \min\left(1, 1 + \frac{\gamma-1}{\theta}\right) \ \& \ \delta < \theta\}$.

If we assume that $\gamma$, $\delta$ and $\theta$ (with $\delta < \theta$) are fixed, the combined behavior of Eqs. (S20), (S23), (S26), and (S29) produces a functional dependence of $\beta$ on $\alpha$ composed by three continuous line segments: an initial horizontal plateau (Eq. S29) followed by an increasing (Eq. S26, when $\gamma > 1$) or decreasing (Eq. S23, when $\gamma < 1$) linear function followed by another horizontal plateau (Eq. S20).

**Case A.5:** $\alpha > \beta$, $\alpha > 1$, and $\delta > \theta$

Under these conditions, $\Lambda_2$ and $\Lambda_4$ dominate the behavior of Eq. (S13) for large $S$, yielding

$$Y \approx \frac{c(\alpha-1)b^{\beta-1}}{\alpha-\beta} S^{1+\lambda_2-\lambda_4}$$
$$\approx \frac{c(\alpha-1)b^{\beta-1}}{\alpha-\beta} S^{1-\theta+\beta\theta}. \tag{S30}$$

By comparing the previous result with Eq. (S6), we find

$$\beta = 1 + \frac{\gamma-1}{\theta} \qquad \text{for } \{\alpha > \beta \ \& \ \alpha > 1 \ \& \ \delta > \theta\}, \tag{S31}$$

and after eliminating $\beta$ from the condition, we have

$$\beta = 1 + \frac{\gamma-1}{\theta} \qquad \text{for } \left\{\alpha > \max\left(1, 1+\frac{\gamma-1}{\theta}\right) \ \& \ \delta > \theta\right\}. \tag{S32}$$

Equation (S32) thus predicts that $\beta$ is a constant when $\{\alpha > \max\left(1, 1+\frac{\gamma-1}{\theta}\right) \ \& \ \delta > \theta\}$.

**Case A.6:** $\alpha > \beta$, $\alpha < 1$, and $\delta > \theta$

For this case, $\Lambda_2$ and $\Lambda_3$ dominate the behavior of Eq. (S13) for large $S$ and produce

$$Y \approx \frac{c(1-\alpha)a^{\alpha-1}b^{\beta-\alpha}}{\alpha-\beta} S^{1+\lambda_2-\lambda_3}$$
$$\approx \frac{c(1-\alpha)a^{\alpha-1}b^{\beta-\alpha}}{\alpha-\beta} S^{1-\delta+\beta\theta+\alpha(\delta-\theta)}. \tag{S33}$$

By comparing the previous result with Eq. (S6), we find

$$\beta = \frac{\gamma+\delta-1}{\theta} + \left(1 - \frac{\delta}{\theta}\right)\alpha \qquad \text{for } \{\alpha > \beta \ \& \ \alpha < 1 \ \& \ \delta > \theta\}, \tag{S34}$$

and after eliminating $\beta$ from the condition, we have

$$\beta = \frac{\gamma+\delta-1}{\theta} + \left(1 - \frac{\delta}{\theta}\right)\alpha \qquad \text{for } \left\{\gamma < 1 \ \& \ 1+\frac{\gamma-1}{\delta} < \alpha < 1 \ \& \ \delta > \theta\right\}. \tag{S35}$$

Equation (S35) predicts a linear decreasing correspondence between $\beta$ and $\alpha$ when $\{\gamma < 1 \ \& \ 1+\frac{\gamma-1}{\delta} < \alpha < 1 \ \& \ \delta > \theta\}$.

**Case A.7:** $\alpha < \beta$, $\alpha > 1$, and $\delta > \theta$

For this case, $\Lambda_1$ and $\Lambda_4$ dominate the behavior of Eq. (S13), yielding

$$Y \approx \frac{c(\alpha-1)a^{\beta-\alpha}b^{\alpha-1}}{\beta-\alpha}S^{1+\lambda_1-\lambda_4}$$
$$\approx \frac{c(\alpha-1)a^{\beta-\alpha}b^{\alpha-1}}{\beta-\alpha}S^{1-\theta+\beta\delta+\alpha(\theta-\delta)}. \tag{S36}$$

By comparing the previous result with Eq. (S6), we find

$$\beta = \frac{\gamma+\theta-1}{\delta} + \left(1-\frac{\theta}{\delta}\right)\alpha \qquad \text{for } \{\alpha < \beta \ \& \ \alpha > 1 \ \& \ \delta > \theta\}, \tag{S37}$$

and after eliminating $\beta$ from the condition, we have

$$\beta = \frac{\gamma+\theta-1}{\delta} + \left(1-\frac{\theta}{\delta}\right)\alpha \qquad \text{for } \left\{\gamma > 1 \ \& \ 1 < \alpha < 1+\frac{\gamma-1}{\theta} \ \& \ \delta > \theta\right\}. \tag{S38}$$

Equation (S38) predicts a linear increasing correspondence between $\beta$ and $\alpha$ when $\{\gamma > 1 \ \& \ 1 < \alpha < 1+\frac{\gamma-1}{\theta} \ \& \ \delta > \theta\}$. Similarly to cases A.2 and A.3, the conditions underlying Eqs. (S35) and (S38) are mutually exclusive, that is, only one of the linear correspondences exists depending on whether $\gamma > 1$ (super-linear country scaling, Eq. S38) or $\gamma < 1$ (sub-linear country scaling, Eq. S35).

**Case A.8:** $\alpha < \beta$, $\alpha < 1$, and $\delta > \theta$

Under these conditions, $\Lambda_1$ and $\Lambda_3$ dominate the behavior of Eq. (S13), yielding

$$Y \approx \frac{c(1-\alpha)a^{\beta-1}}{\beta-\alpha}S^{1+\lambda_1-\lambda_3}$$
$$\approx \frac{c(1-\alpha)a^{\beta-1}}{\beta-\alpha}S^{1-\delta+\beta\delta}. \tag{S39}$$

By comparing the previous result with Eq. (S6), we find

$$\beta = 1 + \frac{\gamma-1}{\delta} \qquad \text{for } \{\alpha < \beta \ \& \ \alpha < 1 \ \& \ \delta > \theta\}, \tag{S40}$$

and after eliminating $\beta$ from the condition, we have

$$\beta = 1 + \frac{\gamma-1}{\delta} \qquad \text{for } \left\{\alpha < \min\left(1, 1+\frac{\gamma-1}{\delta}\right) \ \& \ \delta > \theta\right\}. \tag{S41}$$

21

Equation (S41) predicts that $\beta$ is a constant when $\{\alpha < \min\left(1, 1 + \frac{\gamma-1}{\delta}\right)$ & $\delta > \theta\}$.

Similarly to cases A.1-A.4, the combined behavior of Eqs. (S32), (S35), (S38) and (S41) produces a functional dependence of $\beta$ on $\alpha$ composed by three continuous line segments: an initial horizontal plateau (Eq. S41) followed by an increasing (Eq. S38, when $\gamma > 1$) or decreasing (Eq. S35, when $\gamma < 1$) linear function followed by another horizontal plateau (Eq. S32).

**Case B:** $\alpha = 1$ and $\beta \neq 1$

We now consider the particular case $\alpha = 1$ and $\beta \neq 1$ in order to ensure the continuous behavior of $\beta$ as function of $\alpha$ at $\alpha = 1$. Under these assumptions, the normalization constant in Eq. (S4) is

$$k = \frac{S}{\ln\left(\frac{s_{\max}}{s_{\min}}\right)}, \tag{S42}$$

and by replacing $s_{\min}$ and $s_{\max}$ with Eqs. (S7) and (S8), we find

$$k = \frac{S}{\ln\left(\frac{b}{a}S^{\theta-\delta}\right)}. \tag{S43}$$

The solution of the integral in Eq. (S9) for $\alpha = 1$ is

$$Y = \frac{kc}{\beta - 1}\left(s_{\max}^{\beta-1} - s_{\min}^{\beta-1}\right), \tag{S44}$$

and after replacing $k$ by Eq. (S43), $s_{\min}$ by Eq. (S7), and $s_{\max}$ by Eq. (S8), we find

$$\begin{aligned}Y &= \frac{cS}{\beta - 1}\left(\frac{b^{\beta-1}S^{-\theta+\beta\theta} - a^{\beta-1}S^{-\delta+\beta\delta}}{\ln\left(\frac{b}{a}S^{\theta-\delta}\right)}\right) \\ &= \frac{cS}{\beta - 1}\left(\frac{\Omega_1 - \Omega_2}{\ln\left(\frac{b}{a}S^{\theta-\delta}\right)}\right),\end{aligned} \tag{S45}$$

where

$$\begin{aligned}\Omega_1 &= b^{\beta-1}S^{-\theta+\beta\theta} \\ \Omega_2 &= a^{\beta-1}S^{-\delta+\beta\delta}\end{aligned}. \tag{S46}$$

22

It is worth noticing that Eq. (S45) does not produce a "pure" power-law behavior for large $S$ due to the logarithmic function in its denominator. However, the logarithmic function changes much slower than the power-law functions in the numerator, allowing us to approximate the behavior of Eq. (S45) by a power-law function for large $S$.

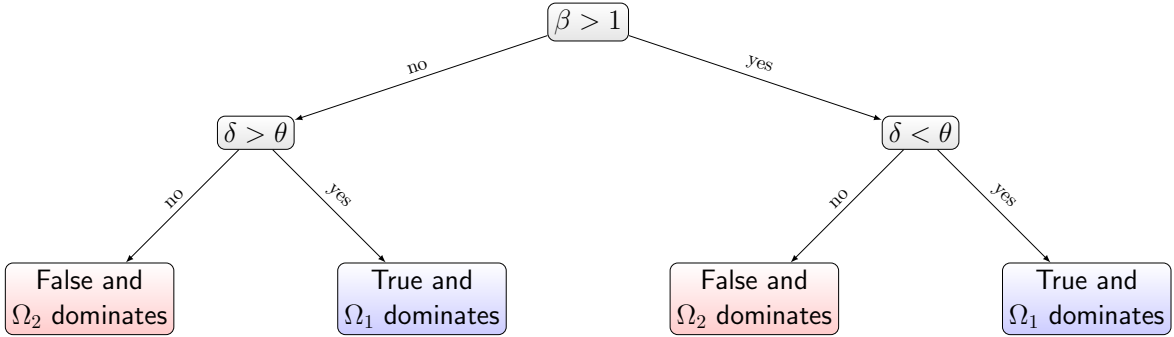

FIG. 14. Decision tree associated with the inequality in Eq. (S47) that defines whether $\Omega_1$ dominates over $\Omega_2$ (or vice-versa) in Eq. (S45).

To estimate the behavior of Eq. (S45) for large values of $S$, we need to determine the conditions under which $\Omega_1$ dominates over $\Omega_2$, that is, when

$$-\theta + \beta\theta > -\delta + \beta\delta$$
$$(\delta - \theta)(1 - \beta) > 0 \,. \tag{S47}$$

The inequality in Eq. (S47) leads us to the decision tree of Figure 14, and from these results, we can find the behavior of $Y$ for large $S$ in Eq. (S45) for each of the following conditions.

**Case B.1:** $\beta > 1$ and $\delta < \theta$

Under these conditions, $\Omega_1$ dominates the behavior of Eq. (S45) for large $S$, leading us to

$$Y = \frac{cb^{\beta-1}}{\beta - 1} \left( \frac{S^{1-\theta+\beta\theta}}{\ln\left(\frac{b}{a}S^{\theta-\delta}\right)} \right) . \tag{S48}$$

By comparing the previous result with Eq. (S6) and considering that $\gamma = 1 - \theta + \beta\theta$, we find

$$\beta = 1 + \frac{\gamma - 1}{\theta} \qquad \text{for } \{\alpha = 1 \,\&\, \beta > 1 \,\&\, \delta < \theta\} \,. \tag{S49}$$

23

This result is the same as obtained for case A.4 and ensures the continuity of the behavior of $\beta$ as a function of $\alpha$ at the point $\alpha = 1$.

**Case B.2:** $\beta < 1$ and $\delta < \theta$

For this case, $\Omega_2$ dominates the behavior of Eq. (S45) for large $S$, leading to
$$Y = \frac{ca^{\beta-1}}{1-\beta}\left(\frac{S^{1-\delta+\beta\delta}}{\ln\left(\frac{b}{a}S^{\theta-\delta}\right)}\right). \tag{S50}$$
By comparing the previous result with Eq. (S6) and considering that $\gamma = 1 - \delta + \beta\delta$, we find
$$\beta = 1 + \frac{\gamma-1}{\delta} \qquad \text{for } \{\alpha = 1 \ \& \ \beta < 1 \ \& \ \delta < \theta\}. \tag{S51}$$
This result is the same as obtained for case A.1 and ensures the continuity of the behavior of $\beta$ as a function of $\alpha$ at the point $\alpha = 1$.

**Case B.3:** $\beta > 1$ and $\delta > \theta$

For this case, $\Omega_2$ dominates the behavior of Eq. (S45) for large $S$, leading us to
$$Y = \frac{-ca^{\beta-1}}{\beta-1}\left(\frac{S^{1-\delta+\beta\delta}}{\ln\left(\frac{b}{a}S^{\theta-\delta}\right)}\right). \tag{S52}$$
It is worth noticing that Eq. (S52) yields positive values for $Y$ since the logarithmic function is negative when $\delta > \theta$ and for large $S$. By comparing the previous result with Eq. (S6) and considering that $\gamma = 1 - \delta + \beta\delta$, we find
$$\beta = 1 + \frac{\gamma-1}{\delta} \qquad \text{for } \{\alpha = 1 \ \& \ \beta > 1 \ \& \ \delta > \theta\}. \tag{S53}$$
This result is the same as obtained for case A.8 and ensures the continuity of the behavior of $\beta$ as a function of $\alpha$ at the point $\alpha = 1$.

**Case B.4:** $\beta < 1$ and $\delta > \theta$

For this case, $\Omega_1$ dominates the behavior of Eq. (S45) for large $S$, leading us to
$$Y = \frac{-cb^{\beta-1}}{1-\beta}\left(\frac{S^{1-\theta+\beta\theta}}{\ln\left(\frac{b}{a}S^{\theta-\delta}\right)}\right). \tag{S54}$$

24

It is worth noticing that Eq. (S54) yields positive values for $Y$ since the logarithmic function is negative when $\delta > \theta$ and for large $S$. By comparing the previous result with Eq. (S6) and considering that $\gamma = 1 - \theta + \beta\theta$, we find

$$\beta = 1 + \frac{\gamma - 1}{\theta} \qquad \text{for } \{\alpha = 1 \ \& \ \beta < 1 \ \& \ \delta > \theta\}. \tag{S55}$$

This result is the same as obtained for case A.5 and ensures the continuity of the behavior of $\beta$ as a function of $\alpha$ at the point $\alpha = 1$.

## 2. FUNCTIONAL DEPENDENCE BETWEEN URBAN SCALING AND ZIPF EXPONENTS

At this point, we can combine the results of cases A1-A8 and write the functional dependence of $\beta$ on $\alpha$ for each of the following conditions.

**For $\gamma > 1$ and $\delta < \theta$:** By combining the cases A.1, A.3 and A.4, we find

$$\beta = \begin{cases} 1 + \frac{\gamma-1}{\theta} & 0 < \alpha \leq 1 \\ \frac{\gamma+\delta-1}{\theta} + \left(1 - \frac{\delta}{\theta}\right)\alpha & 1 < \alpha < 1 + \frac{\gamma-1}{\delta} \\ 1 + \frac{\gamma-1}{\delta} & \alpha \geq 1 + \frac{\gamma-1}{\delta} \end{cases} \quad \text{(S56)}$$

Equation (S56) represents three continuous line segments: an initial horizontal plateau ($\alpha \leq 1$) followed by an increasing linear function ($1 < \alpha < 1 + \frac{\gamma-1}{\delta}$) followed by another horizontal plateau ($\alpha \geq 1 + \frac{\gamma-1}{\delta}$) higher than the first one. Figure 15 illustrates the typical behavior of Eq. (S56).

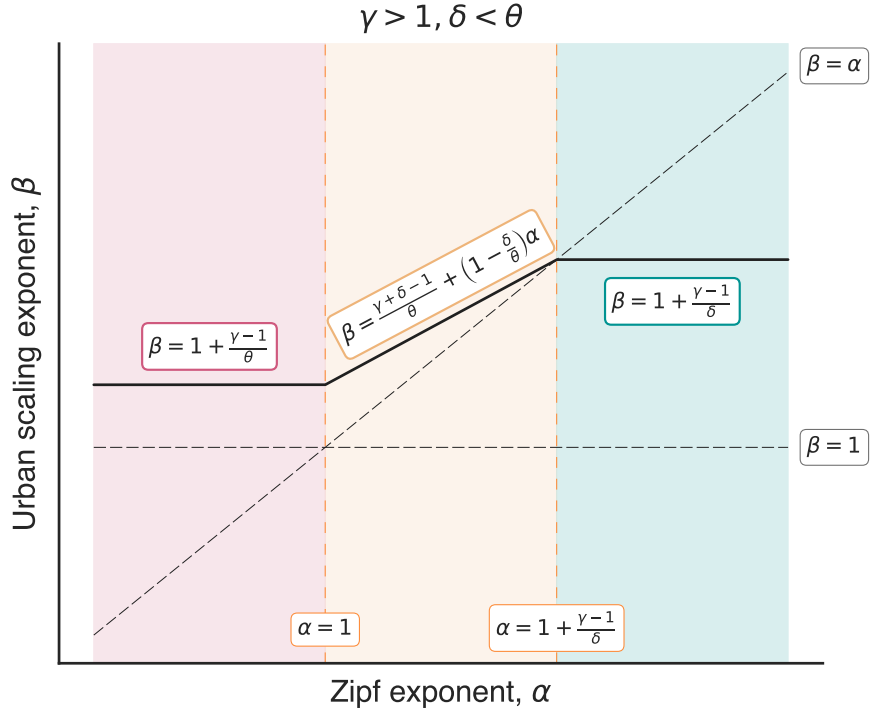

FIG. 15. Illustration of the relation between $\beta$ and $\alpha$ when $\gamma > 1$ and $\delta < \theta$ predicted by Eq. (S56).

**For $\gamma > 1$ and $\delta > \theta$:** By combining the cases A.5, A.7 and A.8, we find

$$\beta = \begin{cases} 1 + \frac{\gamma-1}{\delta} & 0 < \alpha \leq 1 \\ \frac{\gamma+\theta-1}{\delta} + \left(1 - \frac{\theta}{\delta}\right)\alpha & 1 < \alpha < 1 + \frac{\gamma-1}{\theta} \\ 1 + \frac{\gamma-1}{\theta} & \alpha \geq 1 + \frac{\gamma-1}{\theta} \end{cases}. \quad (S57)$$

Equation (S57) represents three continuous line segments: an initial horizontal plateau ($\alpha \leq 1$) followed by an increasing linear function ($1 < \alpha < 1 + \frac{\gamma-1}{\theta}$) followed by another horizontal plateau ($\alpha \geq 1 + \frac{\gamma-1}{\theta}$) higher than the first one. It is worth noticing that we can pass back and forth from Eq. (S56) to Eq. (S57) by replacing $\delta$ by $\theta$.

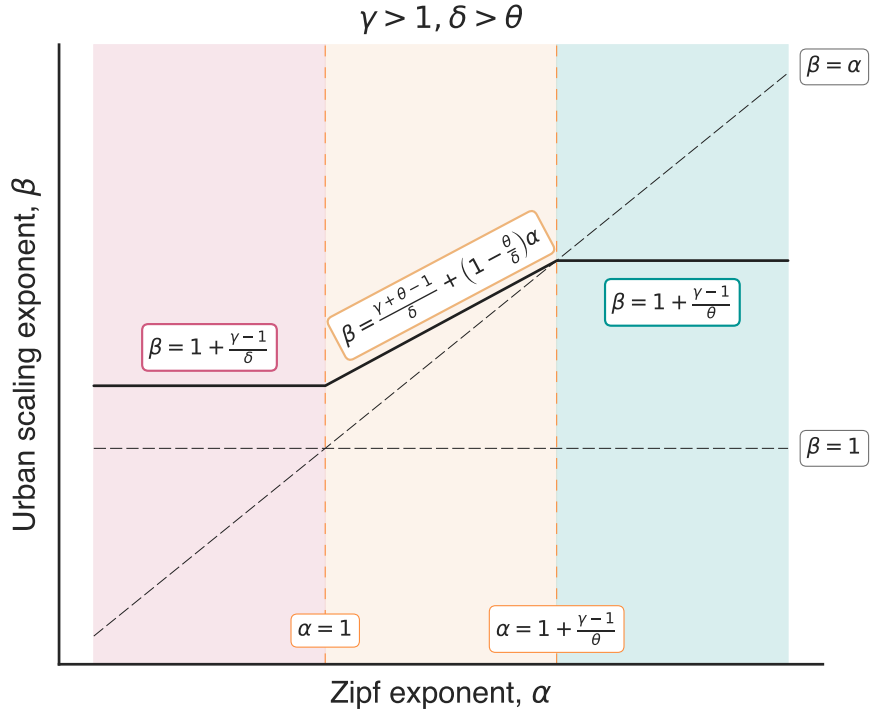

FIG. 16. Illustration of the relation between $\beta$ and $\alpha$ when $\gamma > 1$ and $\delta > \theta$ predicted by Eq. (S57).

**For $\gamma < 1$ and $\delta < \theta$:** By combining the cases A.1, A.2 and A.4, we find

$$\beta = \begin{cases} 1 + \frac{\gamma-1}{\theta} & 0 < \alpha \leq 1 + \frac{\gamma-1}{\theta} \\ \frac{\gamma+\theta-1}{\delta} + \left(1 - \frac{\theta}{\delta}\right)\alpha & 1 + \frac{\gamma-1}{\theta} < \alpha < 1 \\ 1 + \frac{\gamma-1}{\delta} & \alpha \geq 1 \end{cases} \tag{S58}$$

Equation (S58) represents three continuous line segments: an initial horizontal plateau ($\alpha \leq 1 + \frac{\gamma-1}{\theta}$) followed by a decreasing linear function ($1 + \frac{\gamma-1}{\theta} < \alpha < 1$) followed by another horizontal plateau ($\alpha \geq 1$) lower than the first one.

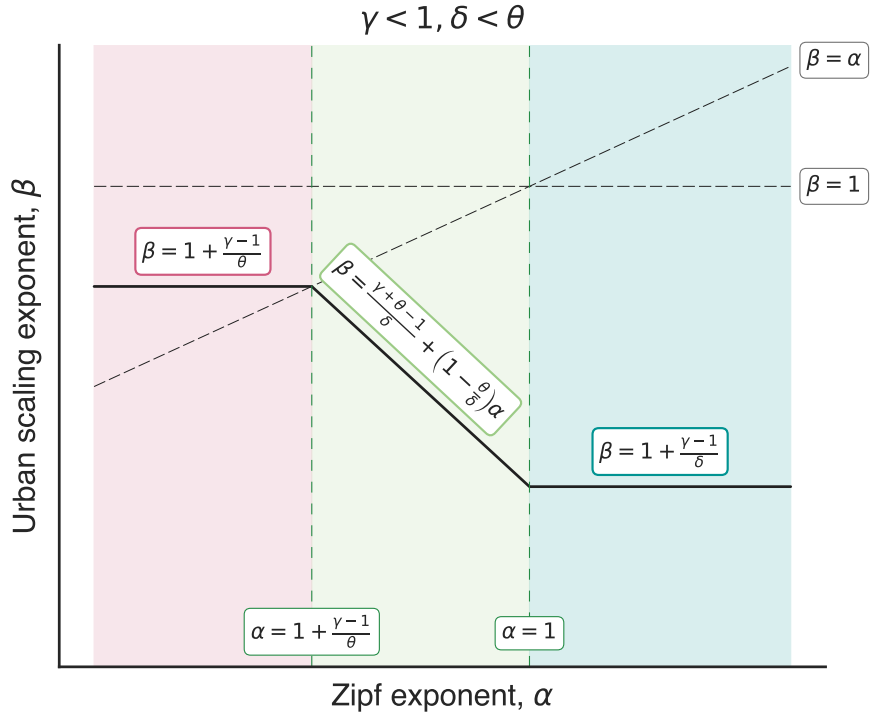

FIG. 17. Illustration of the relation between $\beta$ and $\alpha$ when $\gamma < 1$ and $\delta < \theta$ predicted by Eq. (S58).

**For $\gamma < 1$ and $\delta > \theta$:** By combining the cases A.5, A.6 and A.8, we find

$$\beta = \begin{cases} 1 + \frac{\gamma-1}{\delta} & 0 < \alpha \leq 1 + \frac{\gamma-1}{\delta} \\ \frac{\gamma+\delta-1}{\theta} + \left(1 - \frac{\delta}{\theta}\right)\alpha & 1 + \frac{\gamma-1}{\delta} < \alpha < 1 \\ 1 + \frac{\gamma-1}{\theta} & \alpha \geq 1 \end{cases} \quad (S59)$$

Equation (S59) represents three continuous line segments: an initial horizontal plateau ($\alpha \leq 1 + \frac{\gamma-1}{\delta}$) followed by a decreasing linear function ($1 + \frac{\gamma-1}{\delta} < \alpha < 1$) followed by another horizontal plateau ($\alpha \geq 1$) lower than the first one. It is worth noticing that we can pass back and forth from Eq. (S58) to Eq. (S59) by replacing $\delta$ by $\theta$.

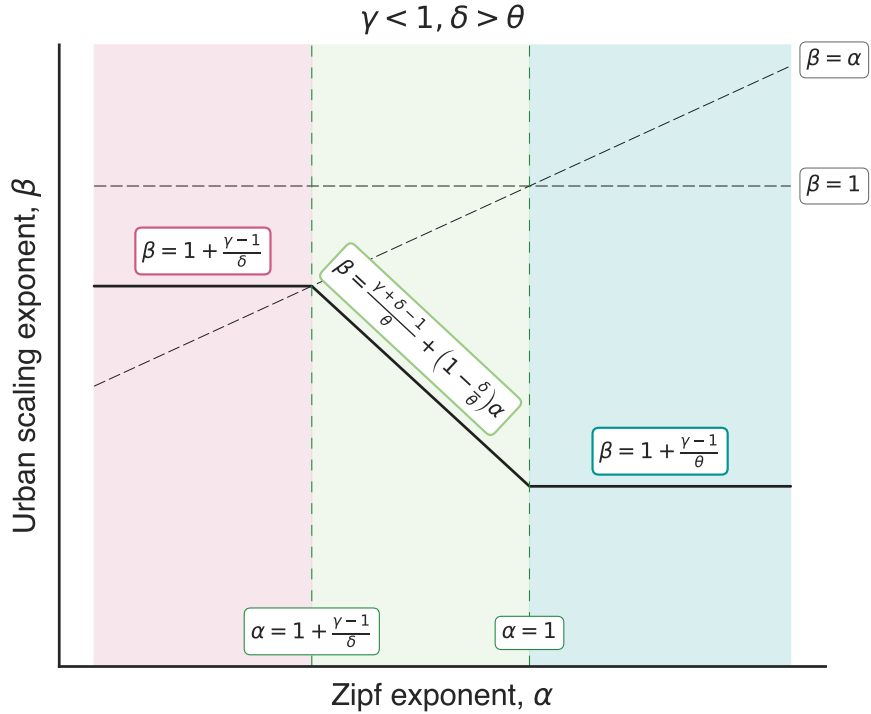

FIG. 18. Illustration of the relation between $\beta$ and $\alpha$ when $\gamma < 1$ and $\delta > \theta$ predicted by Eq. (S59).

**For $\delta = \theta = \varphi$:** By combining the results of Eqs. (S56), (S57), (S58), and (S59), we find

$$\beta = 1 + \frac{\gamma - 1}{\varphi} \qquad \text{for } \alpha > 0, \tag{S60}$$

regardless of $\gamma > 1$ or $\gamma < 1$. Thus, we have that $\beta$ is constant for all values of $\alpha$ when $\delta = \theta$. It is also interesting to note that $\beta = 1$ when $\gamma = 1$, that is, an isometric country scaling implies in isometric urban scaling relationships.

**For $\gamma = 1$:** By combining the results of Eqs. (S56), (S57), (S58), and (S59), we find

$$\beta = 1 \tag{S61}$$

regardless of $\gamma > 1$ or $\gamma < 1$. Thus, constant returns to scale at the country level imply in constant returns to scale at the city level for all values of $\alpha$.

## 3. NUMERICAL SIMULATIONS OF THE CONNECTION BETWEEN URBAN SCALING AND ZIPF EXPONENTS

In order to understand how variations in the country scaling and urban scaling relationships affect the exact expressions of Sec. 2, we have designed a numerical experiment to simulate the empirical relation between $\beta$ and $\alpha$. We begin by generating data at the city level, that is, given the total population $S$, total indicator $Y$, and Zipf exponent $\alpha$ for a country, we generate a list of $m$ population values

$$\mathcal{S} = \{s_1, s_2, \ldots, s_i, \ldots, s_m\}, \tag{S62}$$

where each $s_i$ is a random number drawn from a truncated power-law distribution within the interval $(s_{\max}, s_{\min})$ with exponent $\alpha + 1$ (Eq. S2), and a list of corresponding urban indicators

$$\mathcal{Y} = \{y_1, y_2, \ldots, y_i, \ldots, y_m\}, \tag{S63}$$

where each indicator $y_i$ follows the urban scaling law (Eq. S5) with randon variations

$$\begin{aligned} y_i &= c\, s_i^{\beta(\alpha,\gamma,\delta,\theta)}\, 10^{\mathcal{N}(0,\sigma_y)} \\ \log y_i &= \log c + \beta(\alpha,\gamma,\delta,\theta) \log s_i + \mathcal{N}(0,\sigma_y), \end{aligned} \tag{S64}$$

where $c$ is a constant, $\beta(\alpha,\gamma,\delta,\theta)$ represents the functional relationship between the urban scaling exponent and the Zipf exponent (defined by Eqs. S56-S59) for given values of $\gamma$, $\delta$ and $\theta$), and $\mathcal{N}(0,\sigma_y)$ is Gaussian random variable with zero mean and variance $\sigma_y^2$. The term $\mathcal{N}(0,\sigma_y)$ introduces random variations into the urban scaling relationship, but has a small impact on the country scaling relationships. Thus, in order to account for random variations in the country scaling relationships, we have included analogous terms into Eqs. (S7) and (S8), that is,

$$\begin{aligned} s_{\min} &= a\, S^\delta\, 10^{\mathcal{N}(0,\sigma_\delta)} \\ \log s_{\min} &= \log a + \delta \log S + \mathcal{N}(0,\sigma_\delta) \end{aligned} \tag{S65}$$

and

$$\begin{aligned} s_{\max} &= a\, S^\theta\, 10^{\mathcal{N}(0,\sigma_\theta)} \\ \log s_{\max} &= \log a + \theta \log S + \mathcal{N}(0,\sigma_\theta) \end{aligned}, \tag{S66}$$

where $\mathcal{N}(0, \sigma_\delta)$ and $\mathcal{N}(0, \sigma_\theta)$ are Gaussian random variables with zero mean, and variances $\sigma_\gamma^2$ and $\sigma_\delta^2$, respectively. Finally, we have further considered random variations in the country scaling relationship between $Y$ and $S$ by adding a random term to the total country population, that is,

$$S \to S \, 10^{\mathcal{N}(0,\sigma_\gamma)}, \tag{S67}$$

where $\mathcal{N}(0, \sigma_\gamma)$ is a Gaussian random variable with zero mean and variance $\sigma_\gamma^2$.

In our simulations the values $s_i$ and $y_i$ should satisfy the constraints $\sum_{i=1}^{m} s_i \approx S$ and $\sum_{i=1}^{m} y_i \approx Y$. We satisfy the first constraint by iteratively generating values of $s_i$ and appending to $\mathcal{S}$ until $\sum_{i=1}^{m} s_i \leq S$. To fulfill the second constraint, we define the values of $c$ by comparing Eq. (S6) with results of Case A (Eqs. S18, S21, S24, S27, S30, S33, S36, and S39), that is,

$$c(\alpha, \beta, \delta, \theta, a, b, Y_0) = Y_0 \times \begin{cases} \frac{\alpha-\beta}{(\alpha-1)a^{\beta-1}} & \text{for } \{\alpha > \beta \ \& \ \alpha > 1 \ \& \ \delta < \theta\} \\ \frac{\alpha-\beta}{(1-\alpha)a^{\beta-\alpha}b^{\alpha-1}} & \text{for } \{\alpha > \beta \ \& \ \alpha < 1 \ \& \ \delta < \theta\} \\ \frac{\beta-\alpha}{(\alpha-1)a^{\alpha-1}b^{\beta-\alpha}} & \text{for } \{\alpha < \beta \ \& \ \alpha > 1 \ \& \ \delta < \theta\} \\ \frac{\beta-\alpha}{(1-\alpha)b^{\beta-1}} & \text{for } \{\alpha < \beta \ \& \ \alpha < 1 \ \& \ \delta < \theta\} \\ \frac{\alpha-\beta}{(\alpha-1)b^{\beta-1}} & \text{for } \{\alpha > \beta \ \& \ \alpha > 1 \ \& \ \delta > \theta\} \\ \frac{\alpha-\beta}{(1-\alpha)a^{\alpha-1}b^{\beta-\alpha}} & \text{for } \{\alpha > \beta \ \& \ \alpha < 1 \ \& \ \delta > \theta\} \\ \frac{\beta-\alpha}{(\alpha-1)a^{\beta-\alpha}b^{\alpha-1}} & \text{for } \{\alpha < \beta \ \& \ \alpha > 1 \ \& \ \delta > \theta\} \\ \frac{\beta-\alpha}{(1-\alpha)a^{\beta-1}} & \text{for } \{\alpha < \beta \ \& \ \alpha < 1 \ \& \ \delta > \theta\} \end{cases}, \tag{S68}$$

where the value of $Y_0$ is obtained from the fit of Eq. (S6) to the empirical relationship between $Y$ and $S$ (Fig. 2 of main text).

Algorithm 1 shows the numerical procedures used in our simulations. This algorithm defines the function `generate_country_data` that takes the country population ($S$), the power-law exponents ($\alpha, \gamma, \delta$, and $\theta$), some other constants ($Y_0$, $a$, and $b$), and the parameters related to the intensity of the random variations ($\sigma_y, \sigma_\gamma, \sigma_\delta$, and $\sigma_\theta$) as arguments and returns the simulated lists of populations $\mathcal{S}$ and urban indicators $\mathcal{Y}$. Having these lists, we estimate the simulated exponent $\tilde{\beta}$ in the same way we have proceeded with the empirical data, that is, via robust linear regression of the relationship

**Algorithm 1** Algorithm for generating a list of city populations ($\mathcal{S}$) and a list of urban indicators ($\mathcal{Y}$) for a country with total population $S$.

$(\mathcal{S}, \mathcal{Y}) = \texttt{generate\_country\_data}(S, \alpha, \gamma, \delta, \theta, Y_0, a, b, \sigma_y, \sigma_\gamma, \sigma_\delta, \sigma_\theta)$
    $S = S \, 10^{\mathcal{N}(0,\sigma_\gamma)}$     ▷ $\mathcal{N}(\mu,\sigma)$ is Gaussian random variate with mean $\mu$ and variance $\sigma^2$
    $s_{\min} = a \, S^\delta \, 10^{\mathcal{N}(0,\sigma_\delta)}$
    $s_{\max} = b \, S^\theta \, 10^{\mathcal{N}(0,\sigma_\theta)}$
    $\tilde{S} = 0$
    $\mathcal{S} = []$
    $\mathcal{Y} = []$
    $\beta = \beta(\alpha, \gamma, \delta, \theta)$     ▷ Defined from Eqs. (S56), (S57), (S58), and (S59)
    $c = c(\alpha, \beta, \delta, \theta, a, b, Y_0)$     ▷ Defined from Eq. (S68)
    **while** $\tilde{S} < S$ **do**
        $s_i = \mathcal{P}(\alpha+1, s_{\min}, s_{\max})$     ▷ $\mathcal{P}(\nu, x_{\min}, x_{\max})$ is power-law random variate
                                                                      ▷ with exponent $\nu$ in the interval $(x_{\min}, x_{\max})$
        $y_i = c \, s_i^\beta \, 10^{\mathcal{N}(0,\sigma_y)}$
        $\tilde{S} = \tilde{S} + s_i$
        $\mathcal{S} \longleftarrow s_i$     ▷ Appends $s_i$ to the list of city populations $\mathcal{S}$
        $\mathcal{Y} \longleftarrow y_i$     ▷ Appends $y_i$ to the list of city indicators $\mathcal{Y}$
    **end while**

$\log y_i$ versus $\log s_i$. We also obtain the simulated values for total population $\tilde{S}$ and total indicator $\tilde{Y}$ by summing up the values of $s_i$ and $y_i$. Similarly, we estimate the simulated values of $\tilde{s}_{\min}$ and $\tilde{s}_{\max}$ by taking the maximum and minimum values within the list $\mathcal{S}$.

For our simulations, we use the total population (of all cities within the power-law regime) of each country in our data set as values of $S$. For each of these values of $S$, we also use the corresponding estimated Zipf exponents $\alpha$. In addition, we set up the values $\gamma$, $\delta$, and $\theta$ equal to the best fitting values of the empirical relationship between $\beta$ and $\alpha$. The constants $Y_0$, $a$, and $b$ do not affect the relationship between $\tilde{\alpha}$ and $\tilde{\beta}$, and have been chosen to match the fitted values obtained from the country scaling relationships (with fixed power-laws exponents). We also set the parameters $\sigma_\gamma$, $\sigma_\delta$ and $\sigma_\theta$ equal to the standard deviations of the bootstrap estimates of the empirical exponents $\gamma$, $\delta$, and $\theta$, respectively. Finally, we have varied the parameter $\sigma_y$ and obtained the simulated relationship between $\tilde{\beta}$ and $\tilde{\alpha}$. The same approach is used when considering that $s_i$ follows a power-law distribution with exponential

cutoff, that is,
$$p(s) \sim s^{-(\alpha+1)} \exp(-s/s_c) \tag{S69}$$
where $s_c > 0$ is an additional parameter. The only change necessary to implement this case is to replace the truncated power-law random number $\mathcal{P}(\alpha+1, s_{\min}, s_{\max})$ in Algorithm 1 by an a random number generator associated with Eq. (S69), that is, a $\mathcal{P}(\alpha+1, s_{\min}, s_c)$.

## 4. ZIPF AND URBAN SCALING PLOTS FOR EACH COUNTRY

In this section, we show Zipf's law and urban scaling law adjusted for every country in our data set. In what follows, each row of panels shows the results for a given country (name is indicated within the plots). The first column shows the complementary cumulative distribution $F(s)$ of city population $s$ (gray continuous line) and the adjusted power-law distribution (dashed line). The second column shows the rank plot, where each city is named within the plots. The dashed lines show the adjusted Zipf's law. The power-law exponent $\alpha$ (± standard error) and the lower cutoff population size $s_{\text{smin}}$ (± standard error) are estimated from the approach of Clauset-Shalizi-Newman [SIAM Review 51, 661 (2009)]. The third column shows urban scaling between urban GDP and city population, where each city is also named within the plots. The dashed line represents the adjusted scaling law. The urban scaling exponent $\beta$ (± standard error) is estimated via robust linear regression on the log-transformed quantities.

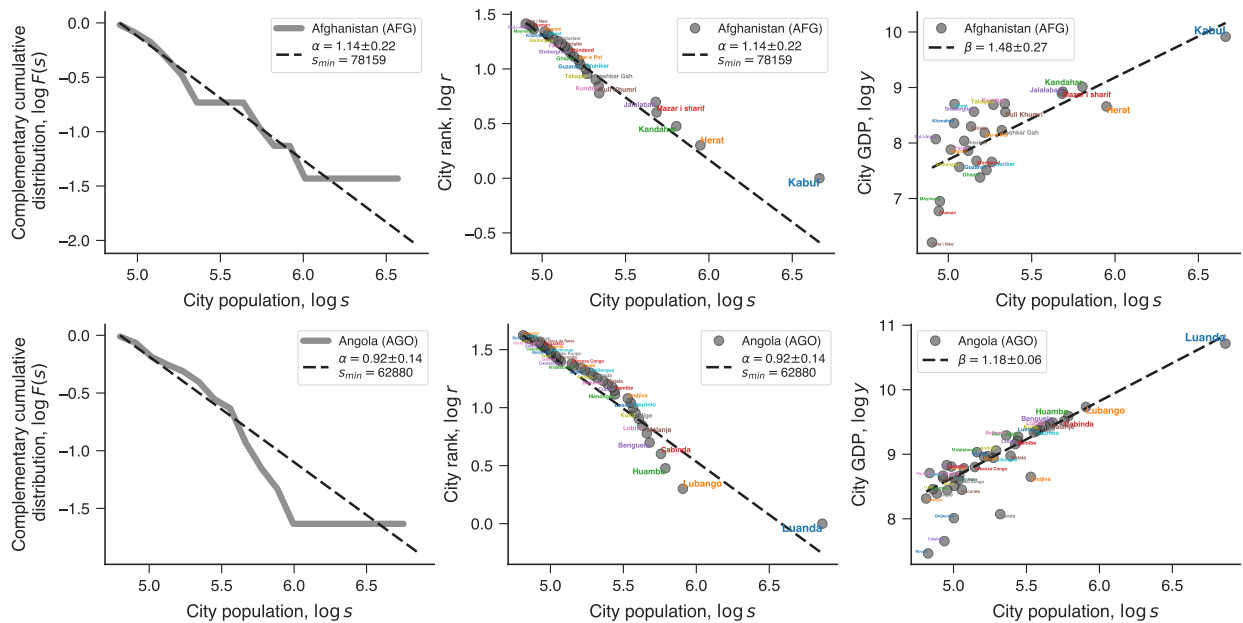

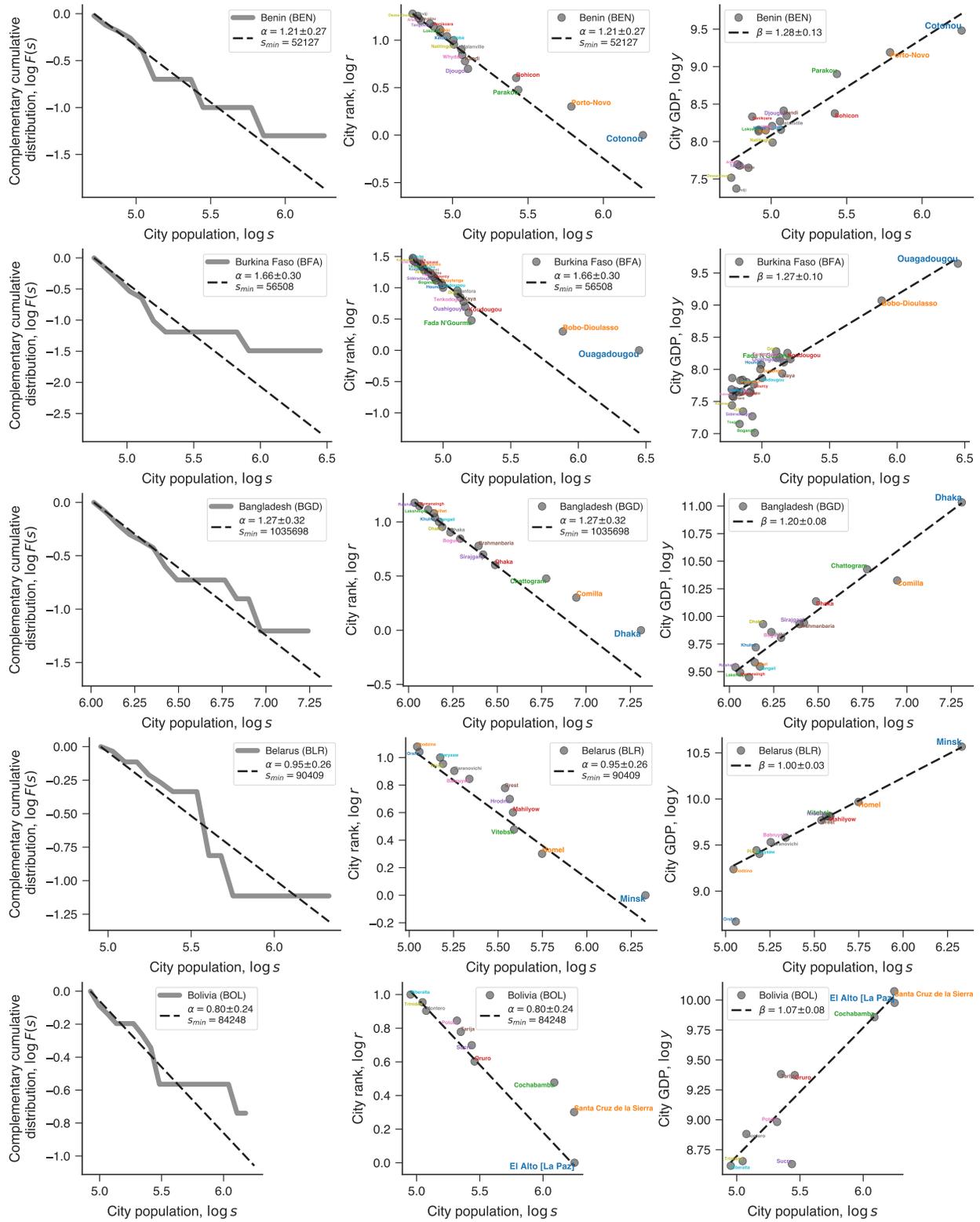

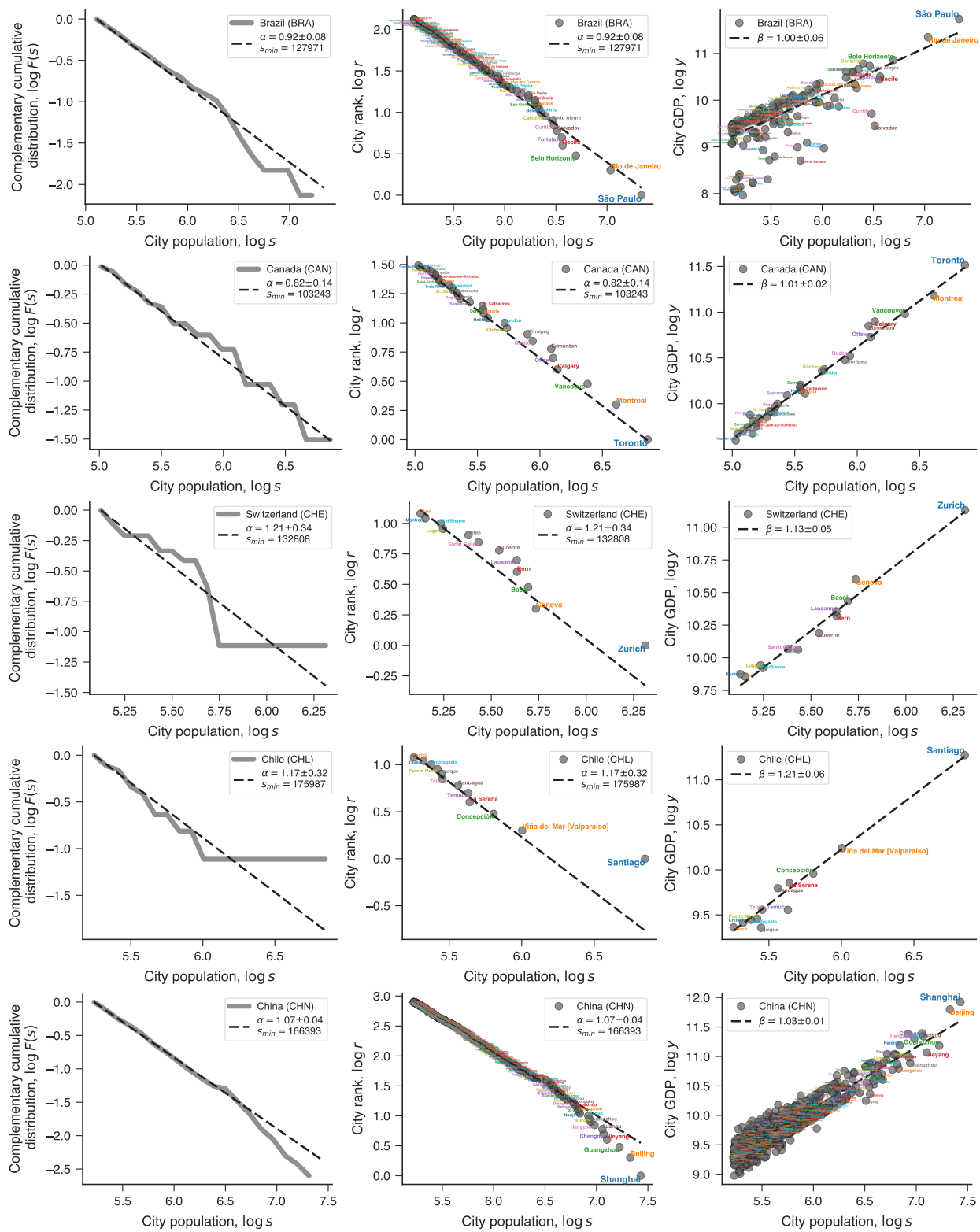

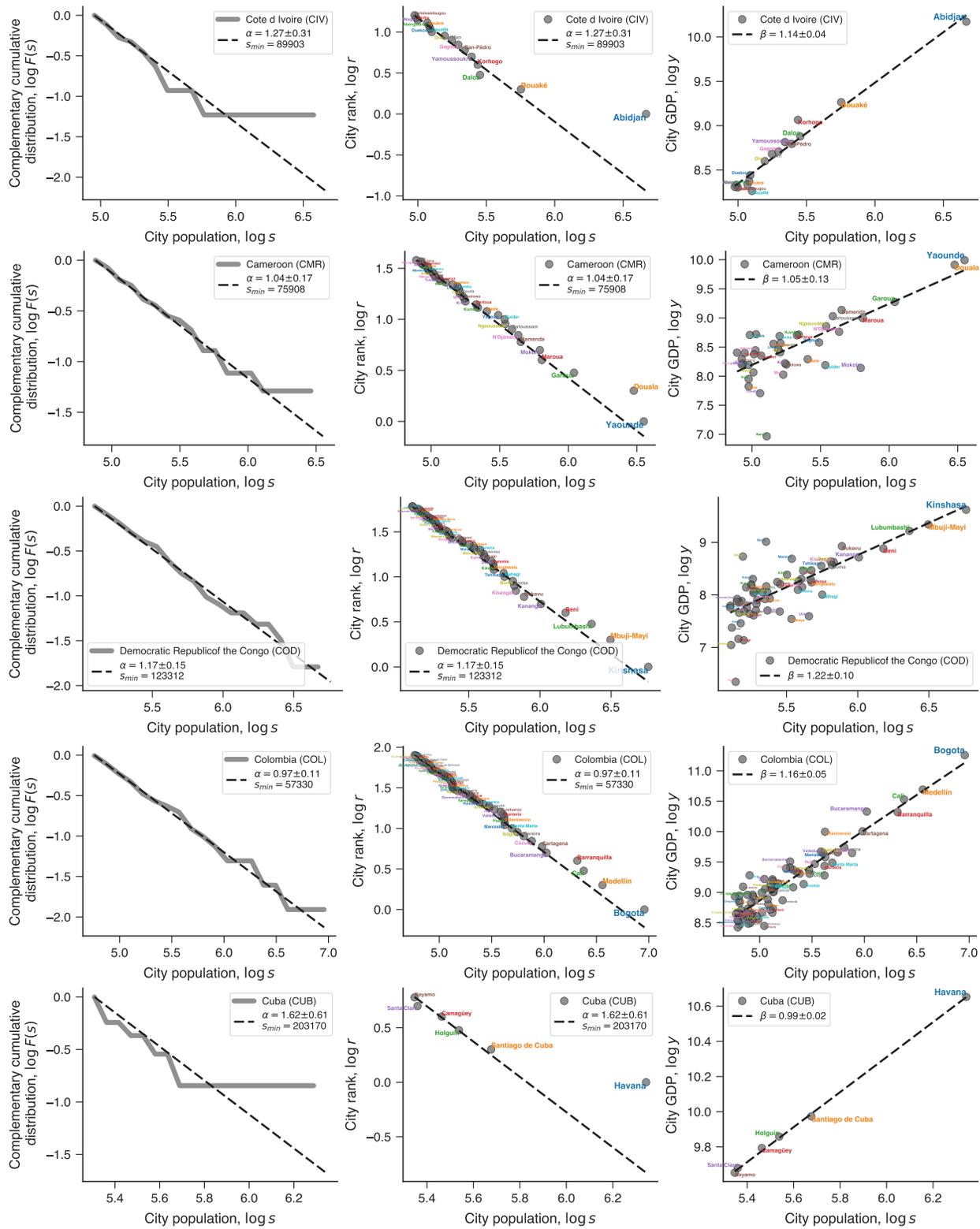

38

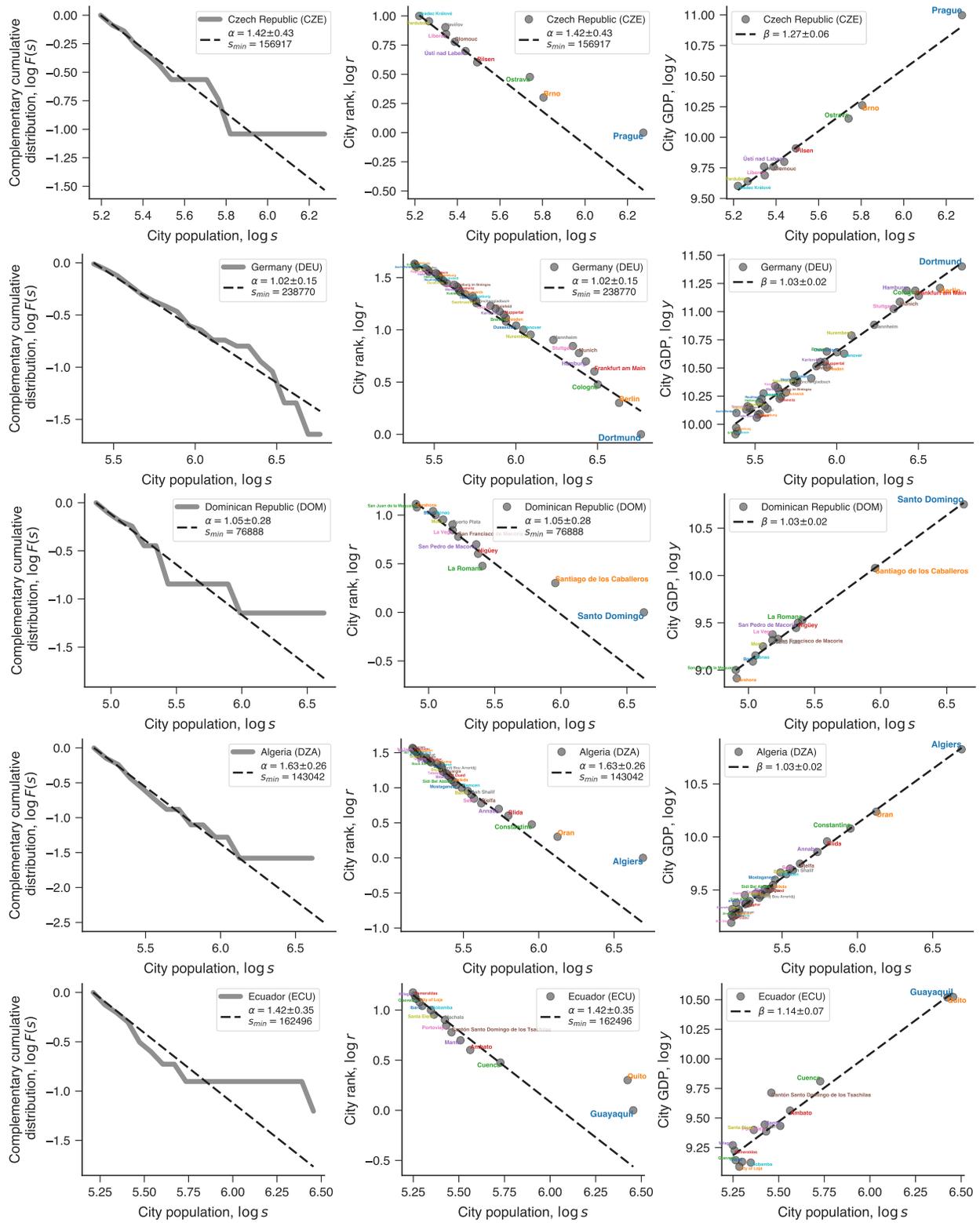
39

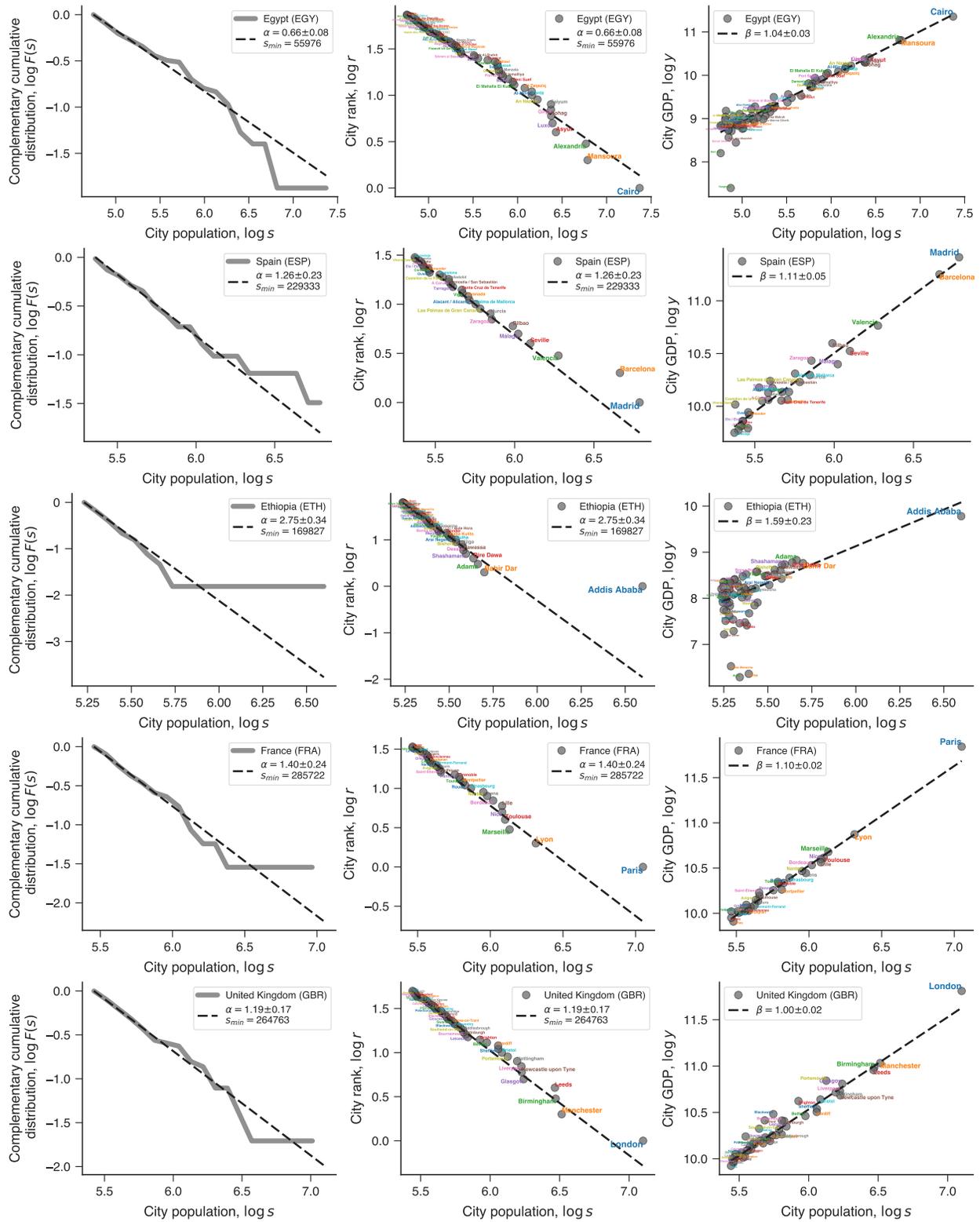



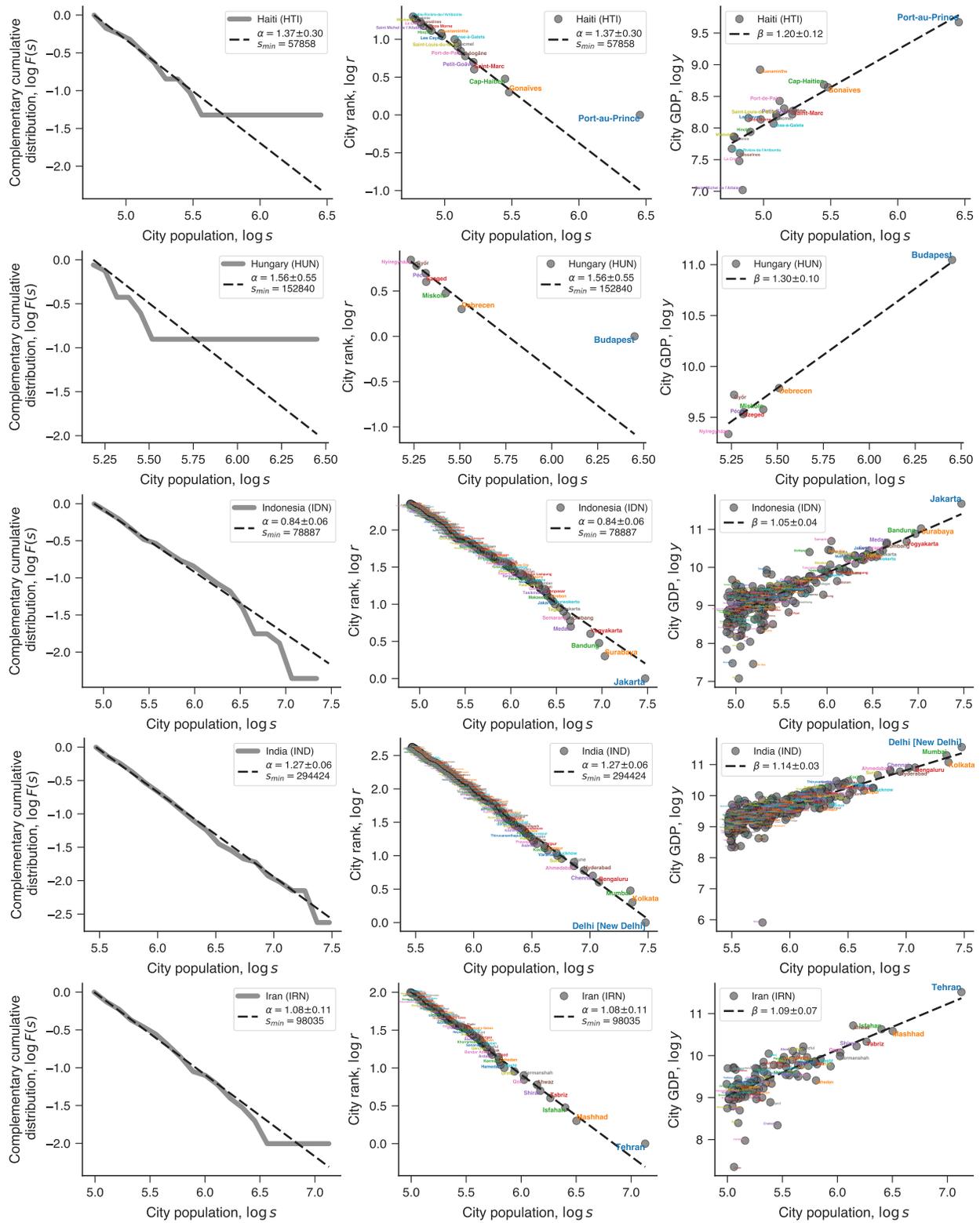

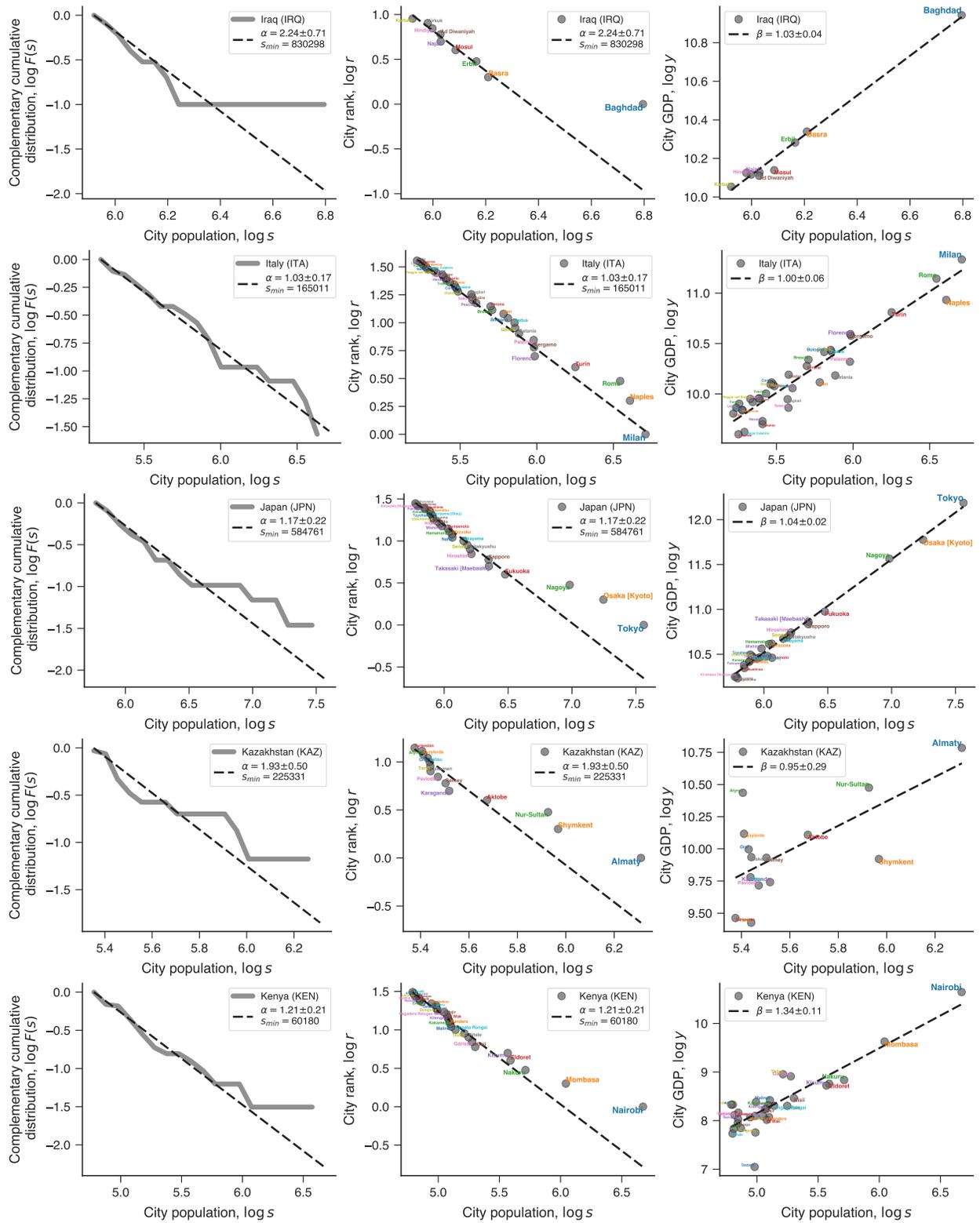

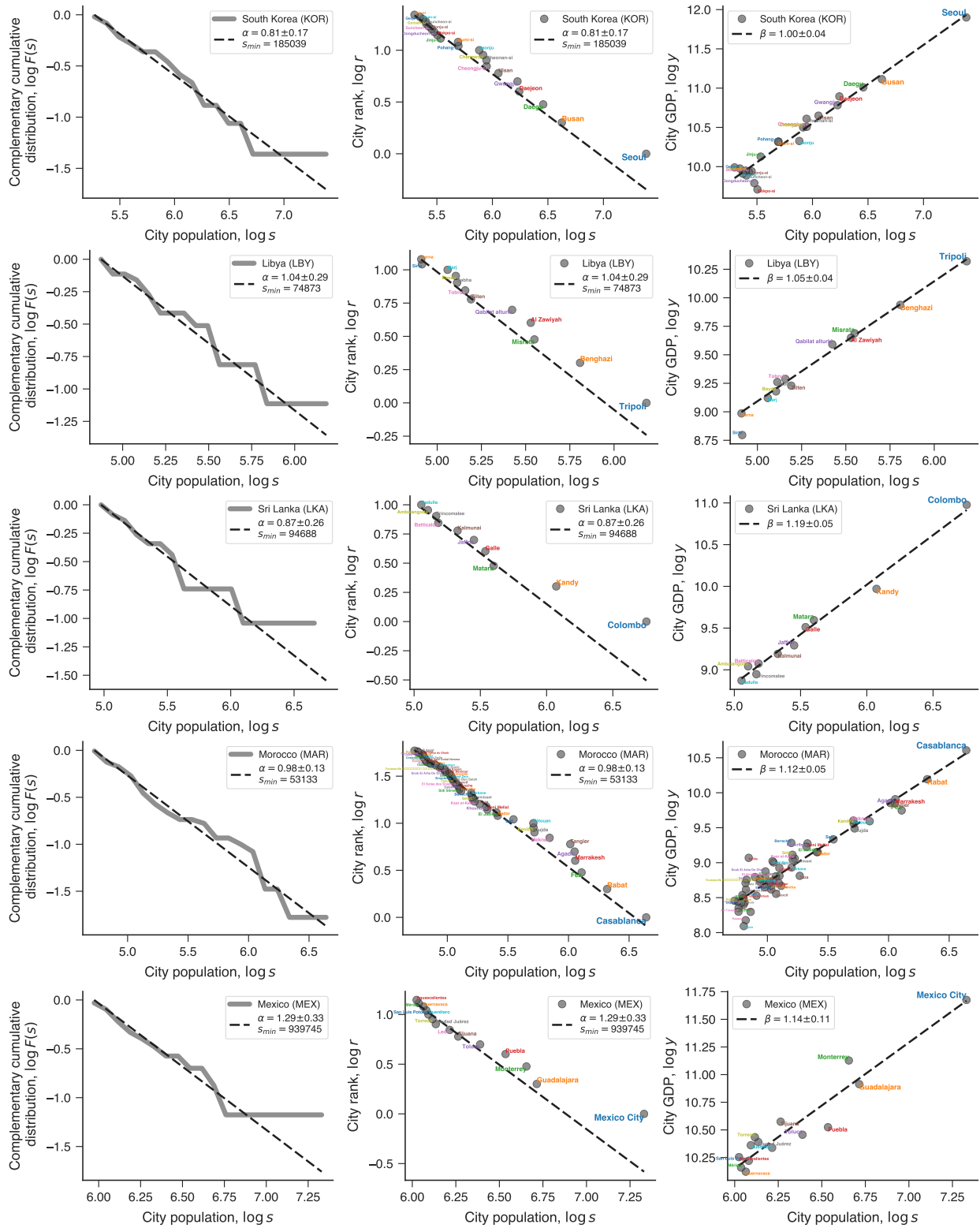

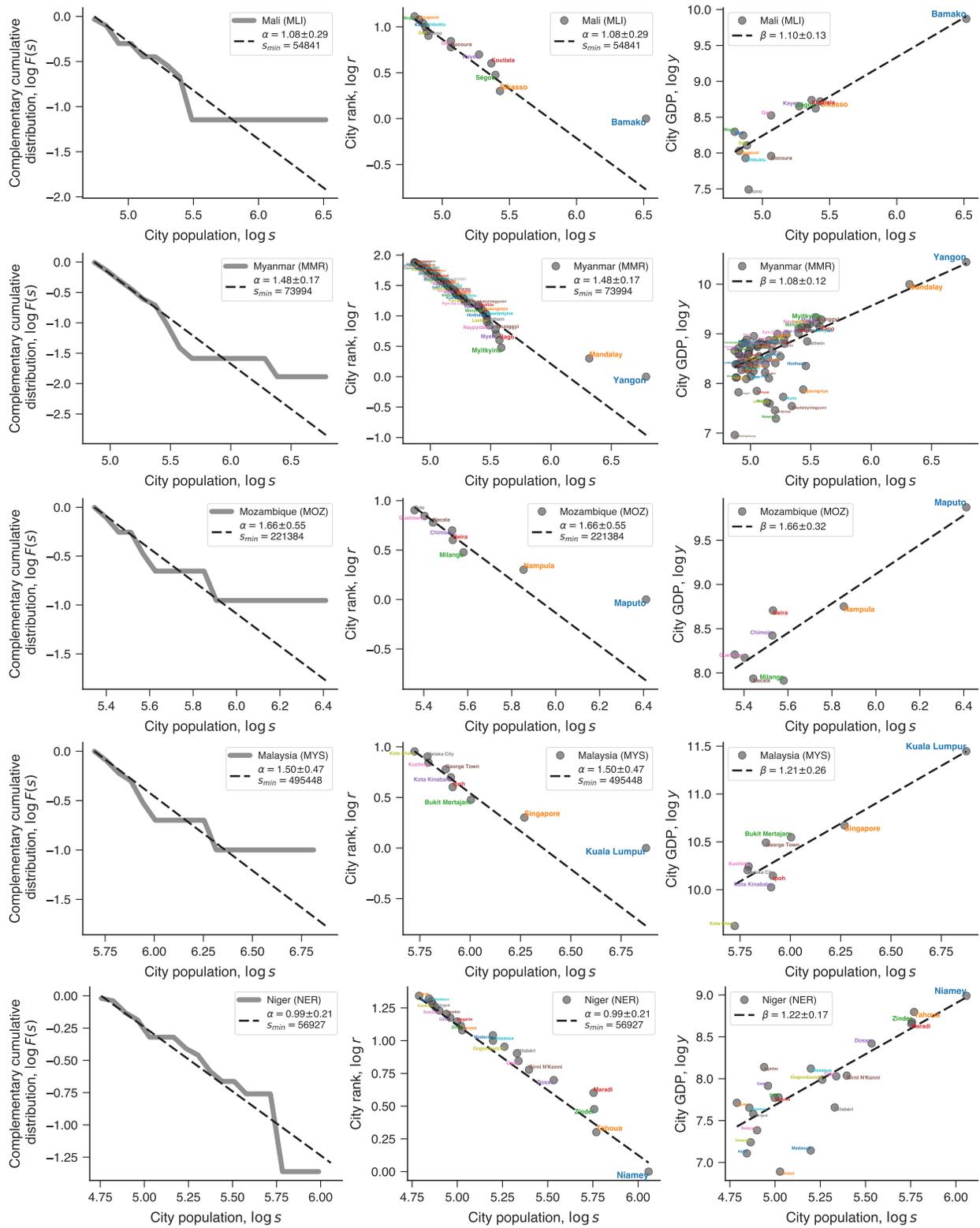

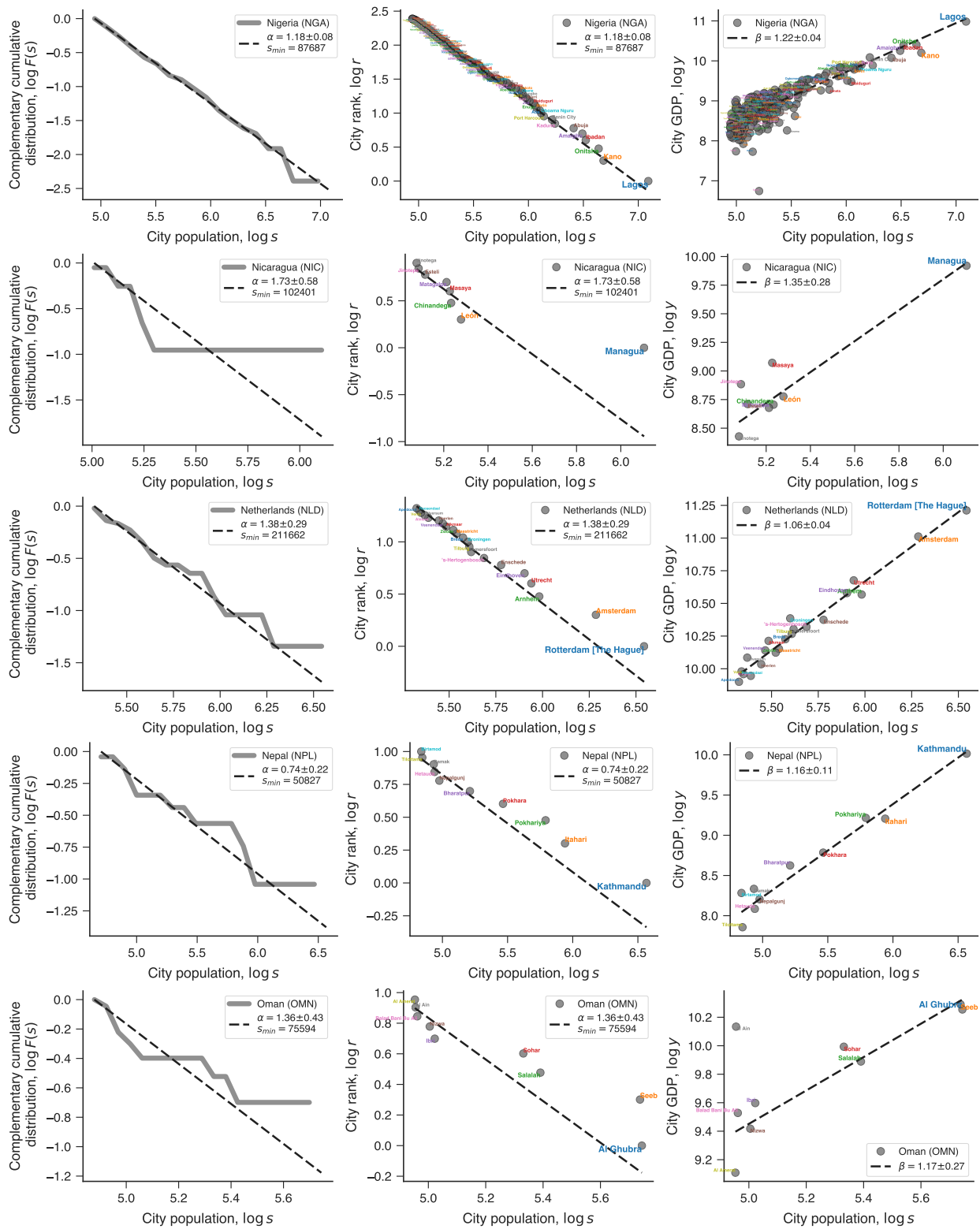

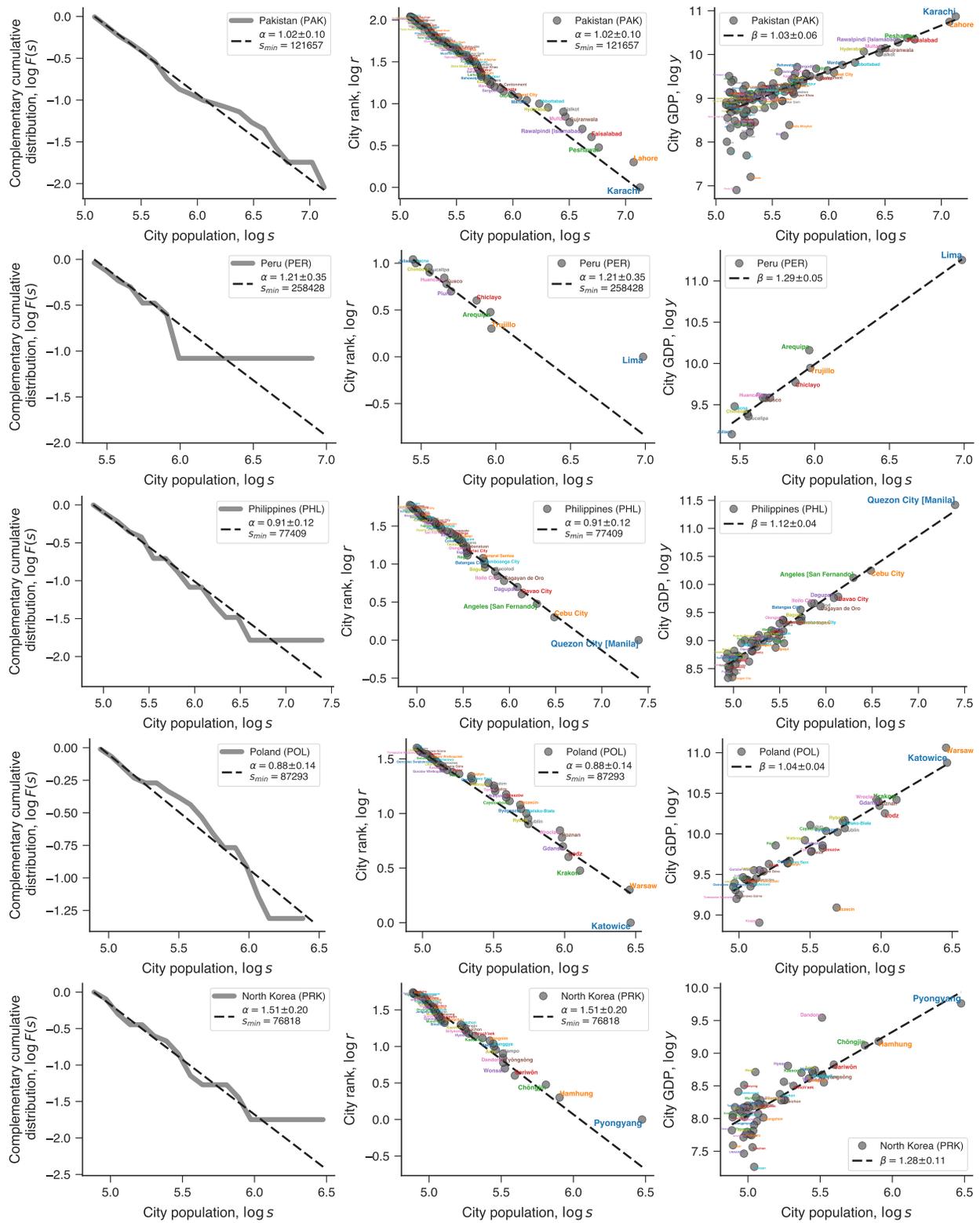

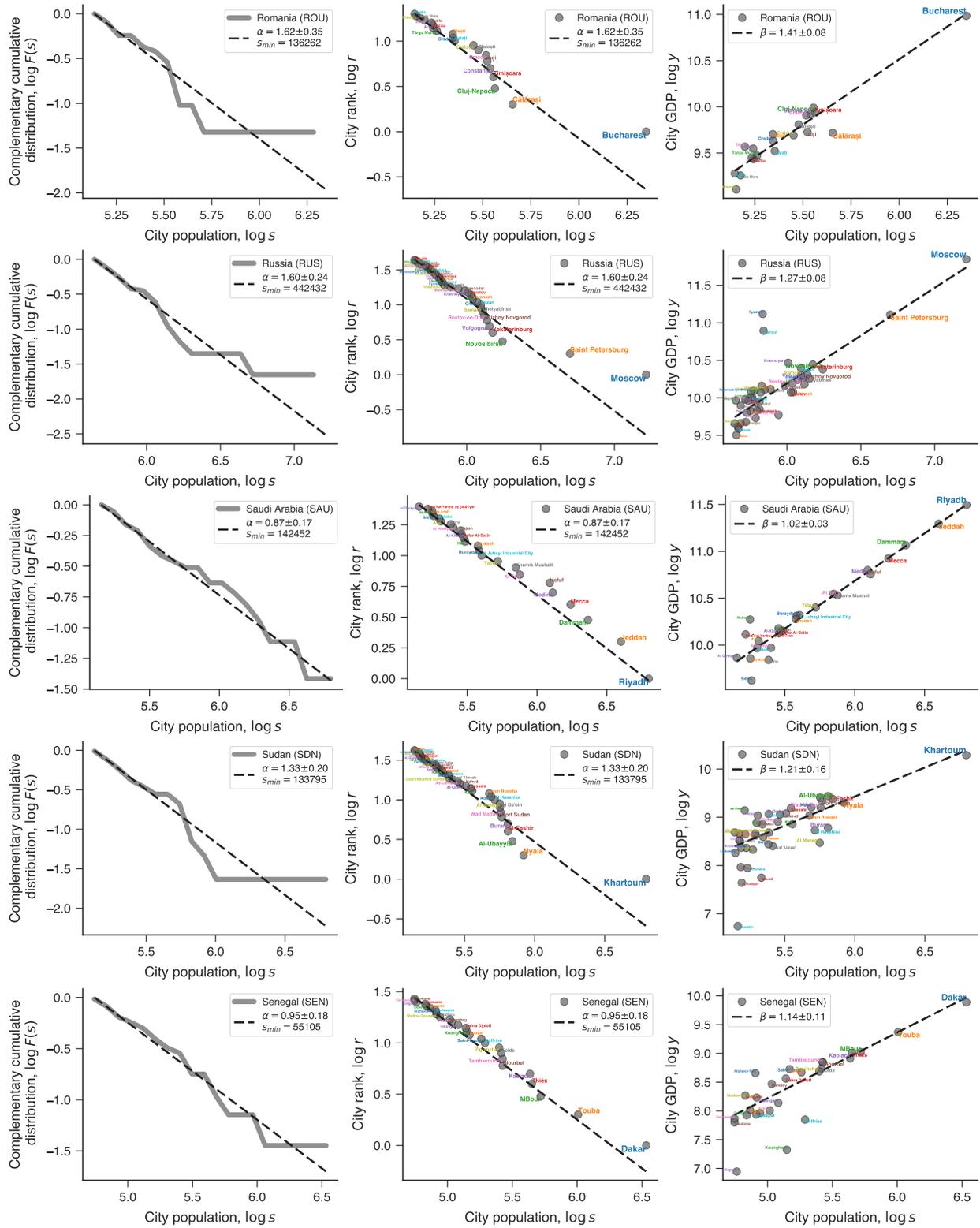

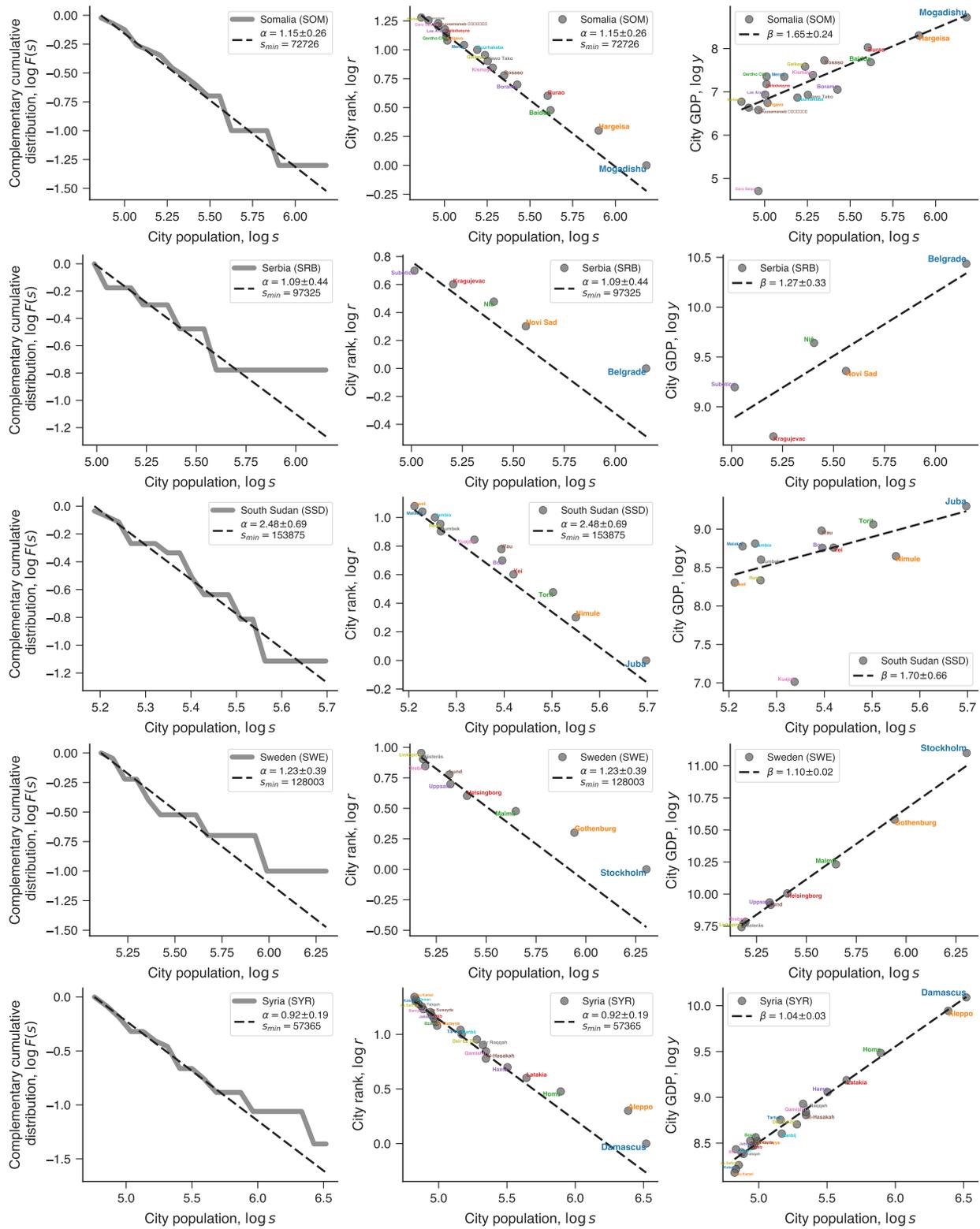



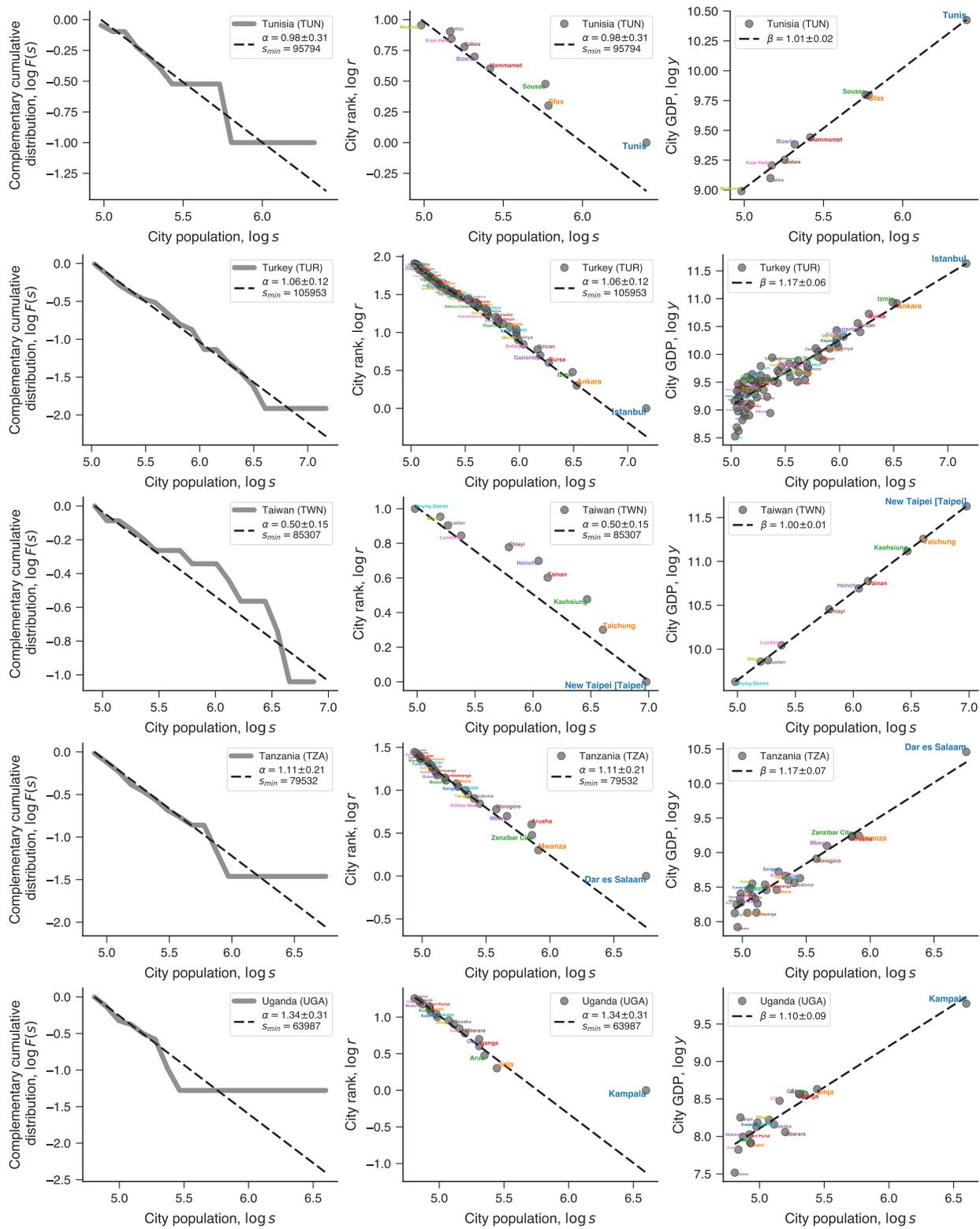

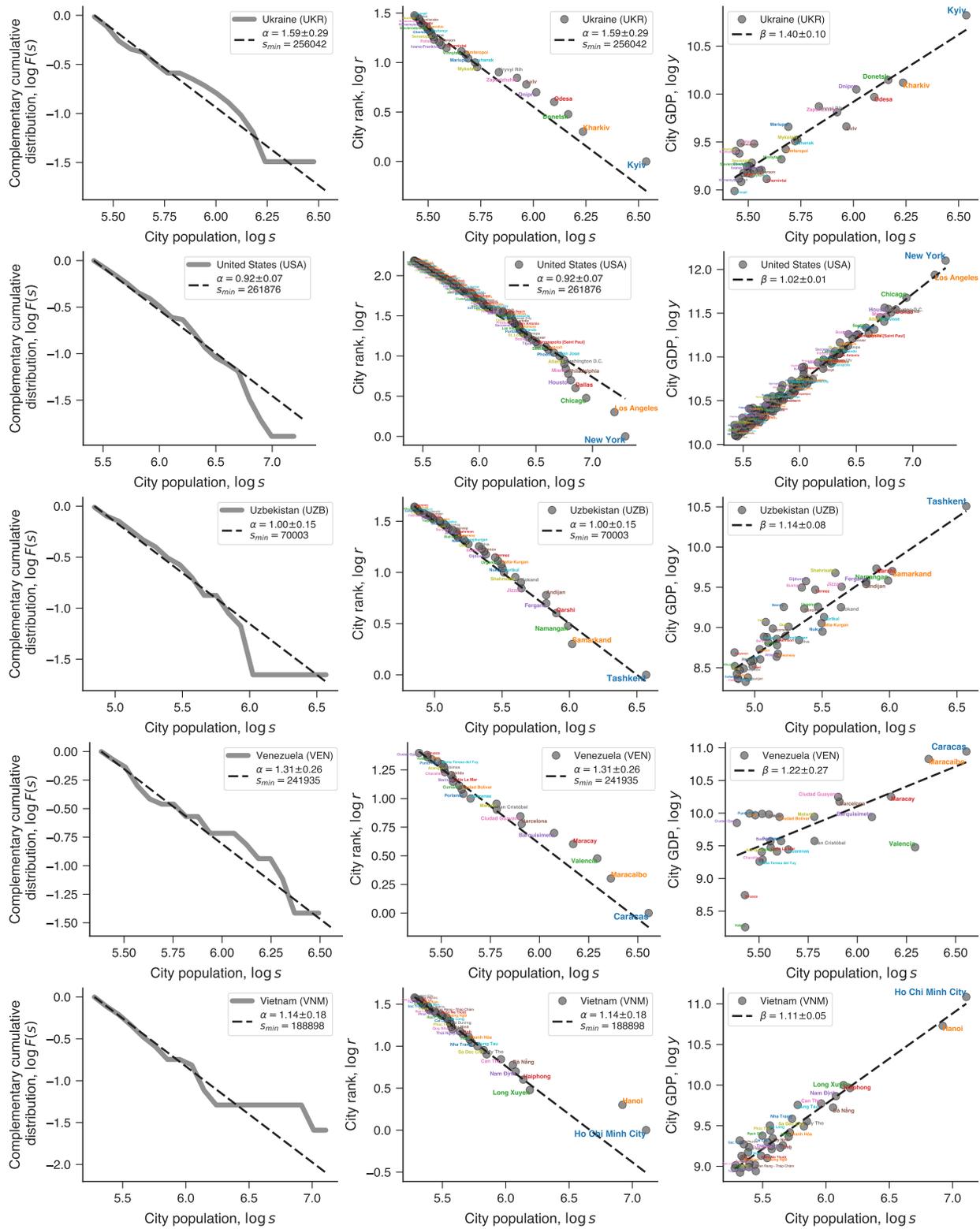

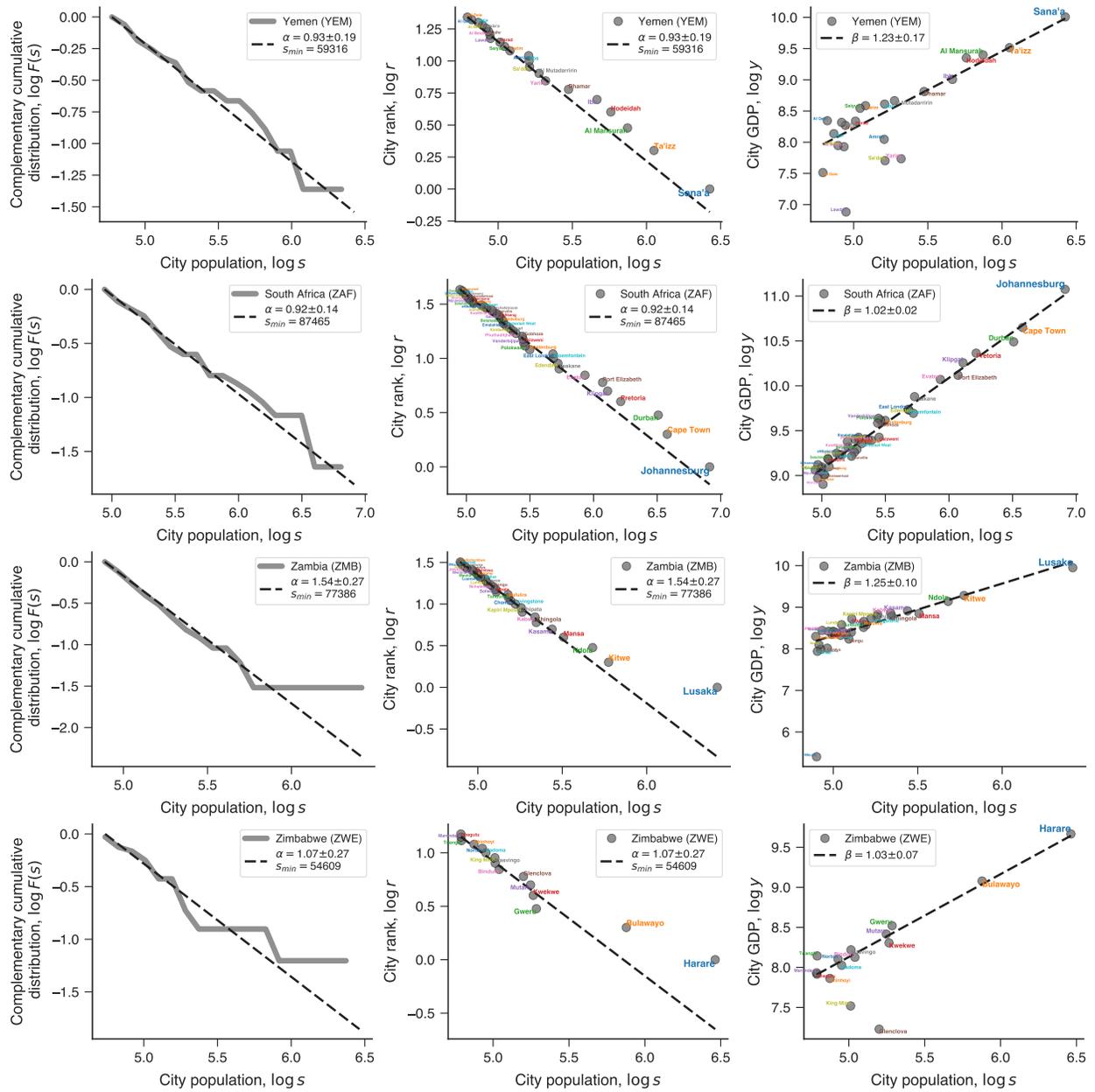



\end{document}